\documentclass[fleqn,10pt]{wlscirep}
\usepackage[utf8]{inputenc}
\usepackage[T1]{fontenc}
\usepackage{bm}
\usepackage{adjustbox}
\usepackage{siunitx}
\usepackage{scalerel}

\newcommand{\hl}[1]{#1}
\newcommand{\boxhl}[1]{#1}
\newcommand{\mathhl}[1]{#1}

\usepackage[mathlines]{lineno}
\usepackage{mathtools}
\usepackage{etoolbox}
\usepackage{float}
\newcommand*\linenomathpatch[1]{%
  \cspreto{#1}{\linenomath}%
  \cspreto{#1*}{\linenomath}%
  \csappto{end#1}{\endlinenomath}%
  \csappto{end#1*}{\endlinenomath}%
}
\newcommand*\linenomathpatchAMS[1]{%
  \cspreto{#1}{\linenomathAMS}%
  \cspreto{#1*}{\linenomathAMS}%
  \csappto{end#1}{\endlinenomath}%
  \csappto{end#1*}{\endlinenomath}%
}
\expandafter\ifx\linenomath\linenomathWithnumbers
  \let\linenomathAMS\linenomathWithnumbers
  \patchcmd\linenomathAMS{\advance\postdisplaypenalty\linenopenalty}{}{}{}
\else
  \let\linenomathAMS\linenomathNonumbers
\fi
\linenomathpatch{equation}
\linenomathpatchAMS{gather}
\linenomathpatchAMS{multline}
\linenomathpatchAMS{align}
\linenomathpatchAMS{alignat}
\linenomathpatchAMS{flalign}

\title{\hl{Inadequacy of fluvial energetics for describing gravity current autosuspension}}
\author[1*]{Sojiro Fukuda}
\author[1]{Marijke G. W. de Vet}
\author[1]{Edward W. G. Skevington}
\author[1]{Elena Bastianon}
\author[1]{Roberto Fern\'andez}
\author[1]{Xuxu Wu}
\author[2]{William D. McCaffrey}
\author[3]{Hajime Naruse}
\author[1]{Daniel R. Parsons}
\author[1]{Robert M. Dorrell}
\affil[1]{Energy and Environment Institute, University of Hull, Hull, United Kingdom}
\affil[2]{School of Earth and Environment, University of Leeds, Leeds, United Kingdom}
\affil[3]{Department of Geology and Mineralogy, Division of Earth and Planetary Sciences, Graduate School of Science, Kyoto University, Kyoto, Japan}

\affil[*]{e-mail: S.Fukuda-2018@hull.ac.uk}

\usepackage{printlen}\uselengthunit{mm}

\begin{document}
\flushbottom
\maketitle
\section*{Abstract}
\begin{quote}
{\centering
    \textit{``Consider the} [turbidity] \textit{current as \hl{...} a river''} R. A. Bagnold (1962);\par}
{\centering
    the foundation of contemporary deep marine sedimentology.\par}
\vspace{2mm}
\hl{Gravity currents, such as sediment-laden turbidity currents, are ubiquitous natural flows that are driven by a density difference. Turbidity currents have provided vital motivation to advance understanding of this class of flows because their enigmatic long run-out and driving mechanisms are not properly understood. 
Extant models assume that material transport by gravity currents is dynamically similar to fluvial flows. 
Here, empirical research from different types of particle-driven gravity currents is integrated with our experimental data, to show that material transport is fundamentally different from fluvial systems. 
Contrary to current theory, buoyancy production is shown to have a non-linear dependence on available flow power, indicating an underestimation of the total kinetic energy lost from the mean flow.
A revised energy budget directly implies that the mixing efficiency of gravity currents is enhanced.}
\end{quote}

\vspace{21pt}
\thispagestyle{empty}

\noindent \textbf{Key points:} turbidity currents, autosuspension, flow power, turbulence


\section*{Introduction}
\hl{
Gravity currents are a broad class of flows with a wide range of environmental applications, including terrestrial cold fronts and submarine thermohaline currents {\cite{Simpson_GCEL}}. 
Of particular interest are particle-driven gravity currents, such as powder snow avalanches, pyroclastic density, and turbidity currents.
Turbidity currents have received significant attention, due to: their capacity to travel long distances, 100s-1,000s of kilometres, along sinuous submarine canyon-channel systems{\cite{MeiburgKneller2010,WellsDorrell2020}}; their importance to the deep marine environment{\cite{WeimerSlatt2007,Picot2019}}; the depositional record of paleoenvironments{\cite{Goldfinger2011}}; and for geohazard risk management\cite{Heezen1952,Bruschi2006,Hsu2008}.
}
\par
\hl{
Turbidity currents are generated by the presence of suspended sediment, meaning they have a higher density than ambient water.
This density difference generates a downslope gravitational force.
The resulting flow produces turbulent mixing, keeping sediment in suspension{\cite{Parker1986}}.
This suspension-flow feedback loop is referred to as autosuspension, the minimal requirement for long runout{\cite{Knapp1938,Bagnold1962}} in all particle-driven gravity currents. 
Accurate prediction of autosuspension is essential to quantify gravity current propagation, natural hazard risk, and, for turbidity currents, deep marine biogeochemical cycling and anthropogenic environmental impact.
However, despite its importance, the mechanisms that enable autosuspension are poorly understood because the kinetic energy of the flow is consumed to maintain the particles in suspension, and uplift them during turbulent mixing with the environment{\cite{WellsDorrell2020}}, which ultimately stalls the flow.
}
\par
\hl{Historically, autosuspension has been explained by the positive feedback whereby sediment entrainment increases the turbulence, referred to as self-acceleration\cite{Parker1986}.}
\hl{Where the slope is steep, such as in the proximal regions of submarine canyons or on volcanic slopes,} gravitational forcing (proportional to the slope) is relatively large \hl{and thus gravity currents} are predominantly net-erosional.
\hl{Entrainment of sediment provides the flow with additional mass and driving force, increasing the momentum and the basal drag, which in turn increases the turbulent energy, further enhancing sediment entrainment and accelerating the flow\cite{Parker1986}.}
\hl{However, turbidity currents can propagate for extensive distances and, in distal reaches, they can traverse near-zero slopes ($< 10^{-4}$ m/m){\cite{Konsoer2013}}.} 
In these regions, the gravitational forcing, proportional to slope, is small.
Consequently\hl{,} the velocity is reduced and the flow is, at best, only weakly erosional, if not net-depositional.
Without net-sediment erosion, the work done due to the entrainment of the ambient fluid\hl{ }decelerates the flow \cite{Parker1986}, and ultimately causes sediment deposition.
The loss of sediment\hl{ reduces} the driving force \hl{and decelerates the flow.}
\hl{This} negative feedback loop stalls the flow, precluding self-acceleration as an explanation of autosuspension.
For partially confined flows, including channel-levee systems, this is exacerbated by the loss of mass and momentum due to \hl{overspill} \cite{Dorrell2014}.
\par
The turbulent kinetic energy (TKE) of a flow is \hl{often taken as} a measure of its capability to suspend sediment\cite{Knapp1938,Bagnold1962,velikanov1954}.
\hl{While there may be local regions where the TKE is dissipated by the flow dynamics, over the total length of the flow the TKE generated and dissipated is vastly greater than the amount stored at any one location.}
\hl{Consequently,} the net advection of TKE is negligible\cite{Caulfield2021}\hl{, and} a necessary condition for autosuspension is that \hl{the energy loss of the mean flow integrated over the bed-normal direction}, \hl{$P_\mathrm{loss}$}, must exceed the \hl{integrated buoyancy production required to maintain the sediment in suspension, $B_\mathrm{gain}$,} and the \hl{integrated} viscous dissipation, $\epsilon$. 
This yields the famed Knapp-Bagnold (K-B) autosuspension criterion\cite{Knapp1938,Bagnold1962}
\begin{equation}
    \mathhl{P_\mathrm{loss}} > \mathhl{B_\mathrm{gain}} +\epsilon.
    \label{eq:tildeP}
\end{equation}
\hl{Bagnold\cite{Bagnold1962} proposed that energy is lost by the mean flow through the shear production of turbulence, and gained by the sediment through turbulent uplift, and this assumption has been widely adopted by subsequent authors\cite{Parker1982,Parker1986,Pratson2000debris,KosticParker2006,Hu2009,PantinFranklin2009,Dorrell2018equilibrium,AmyDorrell2021Equilibrium}. Explicitly, the assumption is that, to leading order,}
\begin{align} \label{eq:turbulent_production}
    \mathhl{P_\mathrm{loss} \simeq P_\mathrm{shear}} &\mathhl{\equiv \int_0^h - \langle u'w' \rangle \frac{\partial \langle u\rangle}{\partial z} {\rm d}z,}
&\text{and}&&
    \mathhl{B_\mathrm{gain} \simeq B_\mathrm{turb}} &\mathhl{\equiv \int_{0}^h gR\langle w^\prime\phi^\prime \rangle\, \textrm{d}z.}
\end{align}
\hl{Here and throughout,} \(\phi\) and \(\Phi\) denote local and depth-averaged volumetric sediment concentration respectively; \hl{$u$ and $U$ denote the local and depth-averaged fluid velocity}; \(g\) denotes gravitational acceleration; \(R=\rho_s/\rho-1\) denotes reduced density; \(w_\mathrm{s}\) denotes particle settling velocity; \(h\) is the extent of the current in the bed normal direction \(z\), i.e. flow depth. Primes and angled brackets denote the Reynolds fluctuations and time-averaged values respectively.
\par
\hl{For equilibrium fluvial flows, $\langle w^\prime\phi^\prime \rangle = \phi{w_{\mathrm{s}}}${\cite{Dorrell2011}}, such that $B_\mathrm{gain} \simeq B_\mathrm{f} \equiv gR\Phi hw_\mathrm{s}$.}
\hl{Moreover, assuming a logarithmic velocity profile, the energy loss in fluvial flow is estimated as $\mathhl{P_\mathrm{loss} \simeq P_{\mathrm{f}}} \equiv u_*^2U$,} where $u_*$ is the shear velocity\cite{Bagnold1962,Parker1986,Pope2006,Andersen2007,AmyDorrell2021Equilibrium}.
\hl{Indeed,} equilibrium \hl{fluvial} flow models for suspended sediment transport, based on the same physical arguments as the K-B criterion \cite{velikanov1954, Bagnold1966approach}, \hl{provide a extensively validated
linear} proportionality between the concentration of suspended sediment and the dimensionless flow power \cite{Wan1994hyper,vanMaren2009,Hu2009,Dorrell2018equilibrium}, 
\begin{align}
    \Phi \propto \frac{\mathhl{P_\mathrm{f}}}{\mathhl{N_\mathrm{f}}} = \frac{u_*^2 U}{gRhw_\mathrm{s}},
    \label{eq:flowpowertheory}
\end{align}
\hl{where $N_\mathrm{f} = B_\mathrm{f}/\Phi$ is the normalised buoyancy production term (the energy required to suspend a unit volume of sediment).}
\par
\hl{A common class of closures for flows are `top-hat' models, where there is no vertical variation in flow structure.
However, via basal shear, these models do capture log-law production energetics to leading order, and the two approaches are equivalent for fluvial systems\cite{Rastogi_Rodi1978OC}.
Such models have been extended to gravity currents{\cite{Ellison1959}}, including turbidity currents{\cite{Parker1986}}, and form the basis of contemporary system scale models\cite{Parker1986,PantinFranklin2009, KosticParker2006,Dorrell2014}. 
The dimensionless flow power inherent to top-hat models of gravity currents can be derived from the kinetic energy conservation equation of the mean flow, which is only modified by the presence of entrainment}
\begin{align}
    \mathhl{
    \frac{P_\mathrm{th}}{N_\mathrm{th}} = \frac{u_*^2 U + \frac{1}{2}e_\mathrm{w}U^3}{gRh\left(w_\mathrm{s}+\frac{1}{2}e_\mathrm{w}U \right)} .
    }
    \label{eq:top-hat-production}
\end{align}
\hl{Here $P_\mathrm{th}$ is derived from the `top-hat' gravity current model as the energy loss of a mean flow, $P_\mathrm{loss}$, and  it is assumed that all energy lost is attributed to the shear production of TKE. 
Moreover, $B_\mathrm{th} = \Phi N_\mathrm{th}=\Phi gRh\left(w_\mathrm{s}+\frac{1}{2}e_\mathrm{w}U \right)$ is the buoyancy production in the model, where $e_\mathrm{w}$ is the water entrainment rate, calculated as the energy required for the sediment to remain uplifted, $B_\mathrm{gain}$, and assumed equal to the turbulent uplift.}
\hl{If there is no entrainment, $P_\mathrm{th} = P_\mathrm{f}$ and $B_\mathrm{th} = B_\mathrm{f}$, thus these are the minimal adjustments to fluvial theory to include entrainment.}
\hl{Consequently, for top-hat models to be valid, it is required that the turbulence in gravity currents is essentially the same as in fluvial systems, despite the substantial differences in the flow structure. 
This implicit assumption is the target of the present analysis.}
\par
Here, the correlation between \hl{the total energy loss of a mean flow, $P_\mathrm{loss}$}, and the \hl{energy required} to keep sediment in suspension, \hl{$B_\mathrm{gain}$}, is reviewed for near-equilibrium \hl{flows}, to investigate whether \hl{idealised} flow power theory is an appropriate predictor for autosuspension\hl{.}
While the \hl{total energy loss of a mean flow} in \hl{gravity} currents\hl{ }is unknown, it has previously been assumed \cite{Bagnold1962,Parker1986,Pope2006,Andersen2007,AmyDorrell2021Equilibrium} proportional to the log-law total-shear TKE production, \hl{$P_\mathrm{f}$} or the \hl{mean-flow energy loss predicted in the top-hat model, $P_\mathrm{th}$}. 
This study reviews experimental and direct observation of \hl{gravity} currents available in the literature, adding \hl{new} experiments to directly address data gaps.
\hl{Crucially, data shows that the total energy loss of the mean flow, $P_\mathrm{th}$, has a non-linear dependence on the work required to keep sediment in suspension, $B_\mathrm{th}$}. 
\hl{A review of the energy deficit implies that particulate transport in gravity currents is driven by mixing at scales larger than that of TKE.}

\begin{figure}[htbp]
\centering
\boxhl{\includegraphics{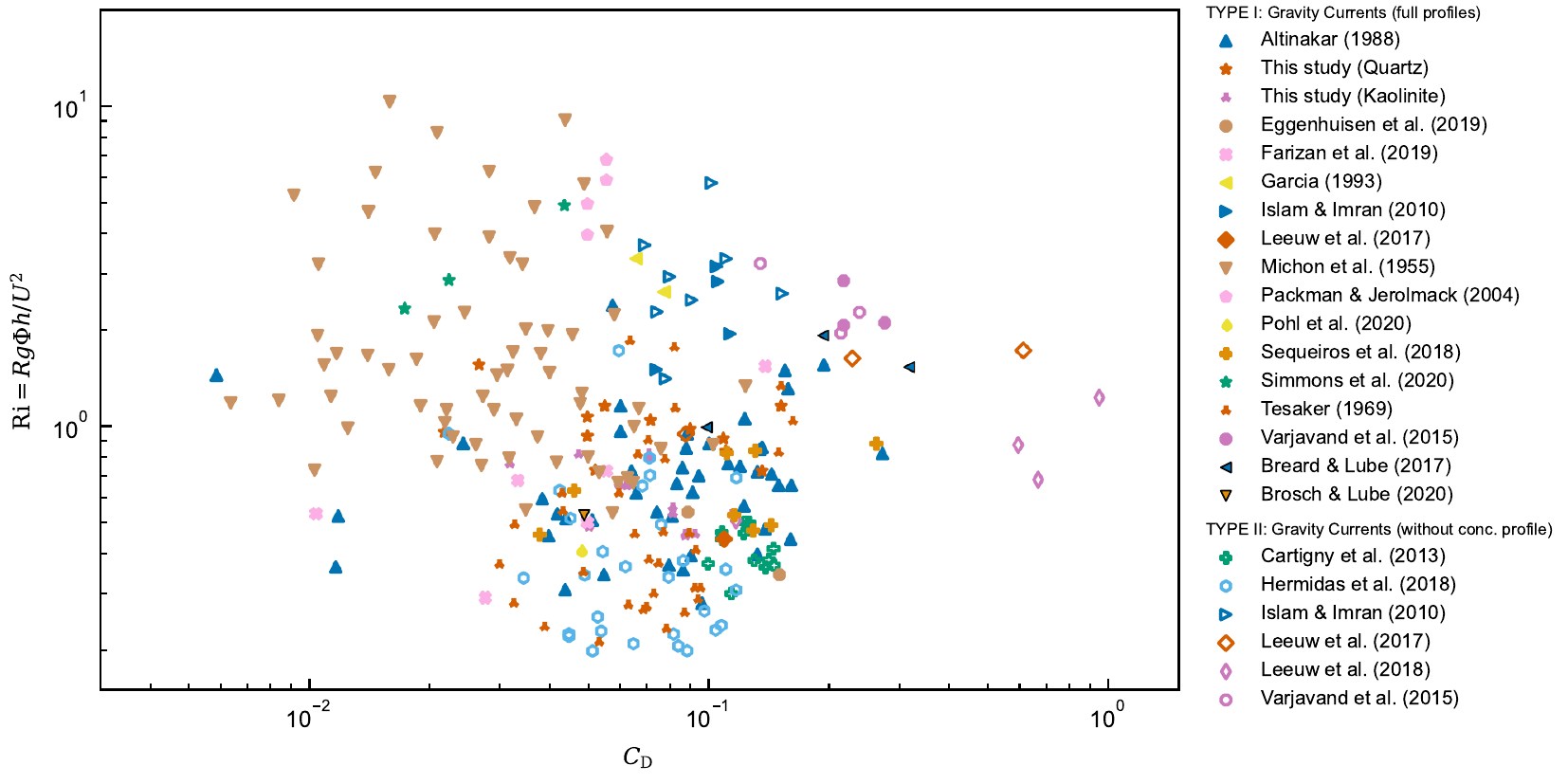}}
\caption{Drag coefficient, $C_\mathrm{D} $\hl{,} and \hl{bulk Richardson number, $\mathrm{Ri}=gR\Phi h/U^2$,} of the compiled \hl{gravity} current data.
Symbols depict field observations and experimental studies of \hl{particle-driven gravity} currents (See \hl{Supplementary Table 1} for the detailed reference and settings of experiments of each literature). \hl{Black borders denoted pyroclastic density current experimental data.}
}
\label{Shields}
\end{figure}

\section*{Results}
\hl{To parameterise energy balance for equilibrium flows in} autosuspension\hl{,} new experiments have been conducted and integrated with over 70 years of empirical and observational data \hl{of turbidity currents and dynamically similar particle-driven pyroclastic density currents}.
\hl{Figure {\ref{Shields}} highlights that this dataset uniformly spans a wide range of flow states, from subcritical to supercritical with small to large drag coefficients}.
Also included are all laboratory-scale studies of constant-discharge sediment-laden turbidity currents in straight channels (see \hl{Supplementary Table 1 and Supplementary Note 2 for more details}).
\hl{The experiments of this study are designed to address extant data gaps, which have limited the full understanding of autosuspension and gravity currents dynamics (see Methods and Supplementary Note 3).}
\hl{Data are separated into types that have both velocity and concentration profiles (TYPE I) and types where concentration is estimated from inlet conditions (TYPE II).}
\par
\begin{figure}[htbp]
\centering
\boxhl{\includegraphics{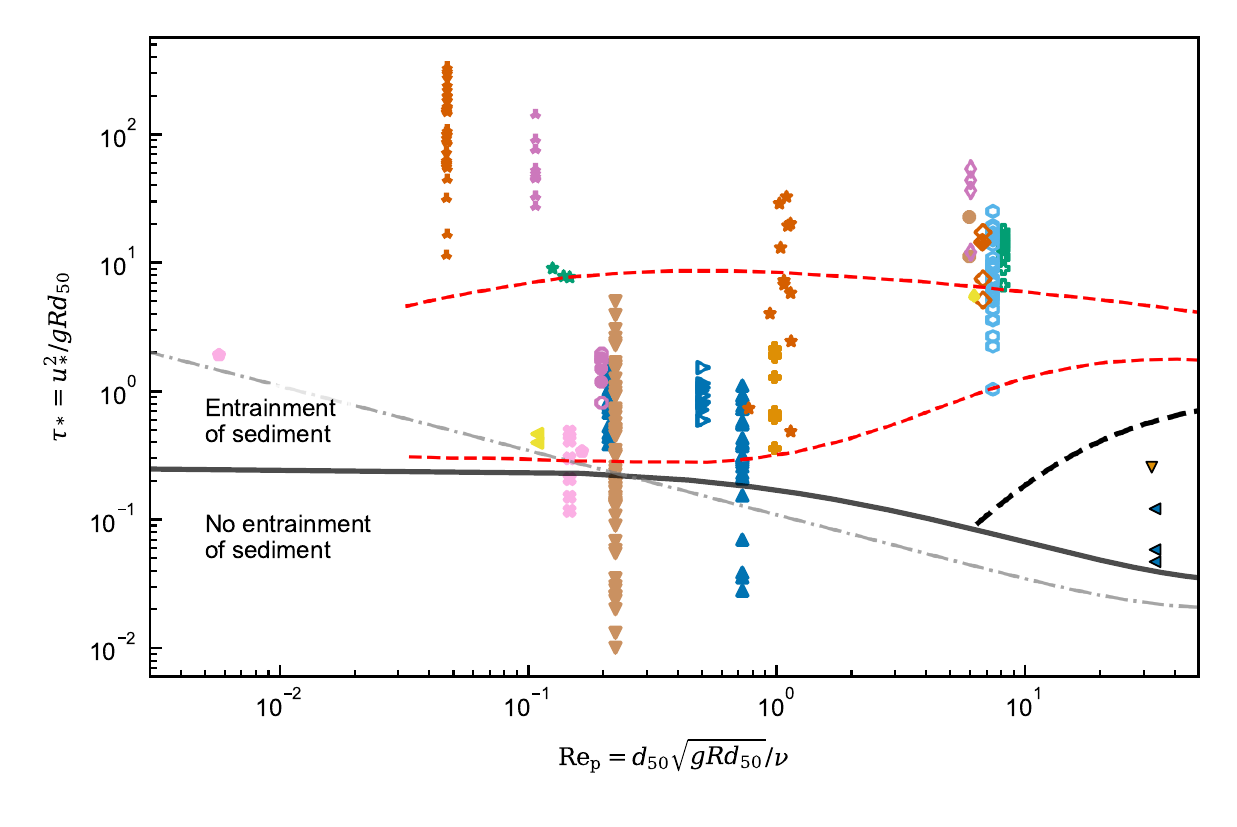}}
\caption{Shields number, $\tau_*$, \hl{versus the} particle Reynolds number, $\mathrm{Re_p}$, of \hl{experimental and observational gravity} current data \hl{(Fig. \ref{Shields})}. 
\hl{Here, $\nu$ denotes the kinematic} fluid viscosity.
Solid\cite{Guo2020} and dash-dot\cite{Parker2003} curves depict the \hl{criteria of incipient motion}.
The black dashed line depicts the $w_\mathrm{s} = u_*$ criterion \hl{for suspended load}\cite{Bagnold1966approach}.
Equilibrium sediment suspension criteria for monodisperse and poorly-sorted suspensions are depicted by the lower and upper red dashed curves respectively \cite{Dorrell2018equilibrium}.
\hl{The symbols and colours are as per Figure {\ref{Shields}}}.
}
\label{DF}
\end{figure}
\hl{The strict requirement for no net sediment deposition is used to limit data parameterising the energy balance of autosuspension.}
When the bed shear stress, described by the dimensionless Shields number, \hl{$\tau_* =u_*^2/gRd_{50}$ where $d_{50}$ is the median particle size,} is less than the threshold needed for incipient sediment motion no sediment can be maintained in suspension: the flow is depositional.
\hl{In Figure {\ref{DF}} the criterion of Guo{\cite{Guo2020}} (solid curve) is taken as the minimal $\tau_*$ for incipient motion.}
\hl{Thus none of the TYPE II data and 19\% of TYPE I data (38 out of 203 points) are excluded as belonging to strictly dispositional flow.}
The remaining turbidity current data lies within the suspended load regime for dilute flows\cite{Bagnold1966approach}.
\par

\paragraph{\hl{Sediment transport} capacity} 
\begin{figure}[htbp]
\boxhl{\includegraphics{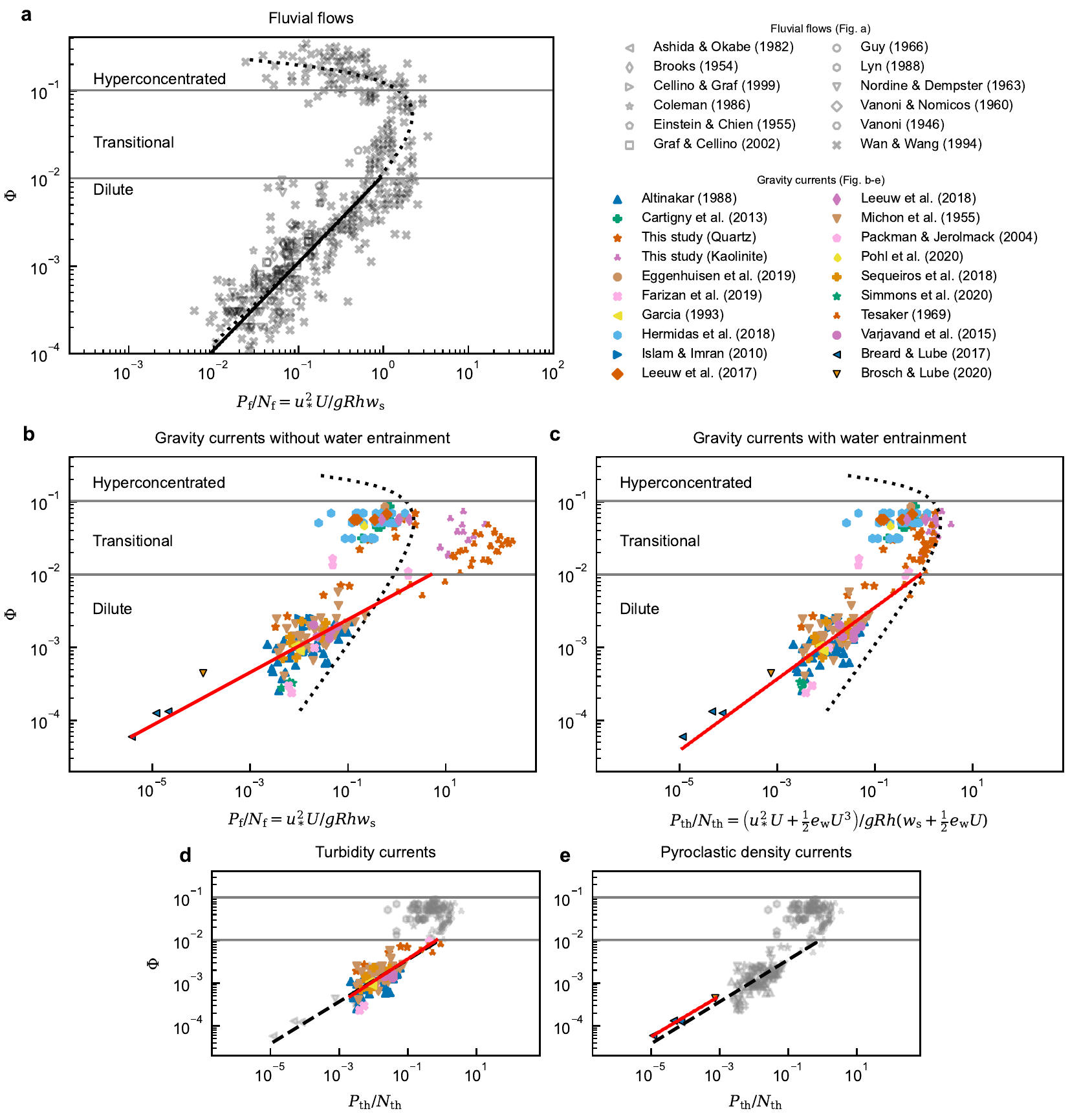}}
\caption{Sediment transport capacity for \hl{fluvial} flows and \hl{gravity} currents.
Sediment concentration, $\Phi$, versus dimensionless flow power, \hl{$P_\sim/N_\sim$ for: 
a) fluvial data (see supplementary material for the detailed reference of each source) with the log-law model, $P_\mathrm{f}/N_\mathrm{f}$;
b) all gravity current data with the log-law model;
c) all gravity current data with the top-hat, entrainment based, model, $P_\mathrm{th}/N_\mathrm{th}$;
d) turbidity current concentration with the top-hat model; and
e) pyroclastic density current concentration with the top-hat model.}
Dilute, transitional and hyperconcentrated regimes are separated by grey solid lines.
\hl{The parametric correlation} of concentration and dimensionless flow power \hl{in fluvial systems} (fitted using an iterative least squares method, see supplementary material) \hl{is} depicted by \hl{a} black dotted curve.
Power-law correlations (Table \ref{tab:fitparams}) in dilute regimes of \hl{each subset} are depicted by black \hl{(a)} and red \hl{(b--e)} solid lines.
\hl{Black dashed lines (d--e) represent the power-law correlation of dilute gravity currents (red solid line in Fig. c).}
}
\label{fpplot}
\end{figure}

In equilibrium flows, \hl{the total kinetic energy loss of the mean flow, $P_\mathrm{loss}$, is assumed to be well approximated by simplified `top-hat' models.
The energy loss balances both the work done to keep sediment in suspension and viscous dissipation, Equation {\eqref{eq:tildeP}}.}
When TKE production is dominated by the effects of basal drag \hl{then the energy available to uplift sediment is given by the log-law of the wall, $P_\mathrm{loss} \simeq P_\mathrm{f}$. Thereby,} a linear correlation is implied between \hl{the} volumetric concentration and dimensionless flow power\hl{, $P_\mathrm{f}/ N_\mathrm{f}$,} where\hl{ $N_\mathrm{f} = gRhw_\mathrm{s}$ }.
This flow-power balance is tested against compiled near-equilibrium laboratory- and natural-scale \hl{gravity} currents (Fig. \ref{fpplot} \hl{and Supplementary Note 4--5}).
\par
\begin{table}[tp]
    \centering 
    \begin{tabular}{l l l c c c}
        \hline
        &Flow type&Subsets&Figure&Curve fit&$R^2$\\ 
        \hline
        &&&&&\\[-1em]
        i)&\hl{Fluvial flows}&\hl{Dilute ($P_\mathrm{f}$, linear fit)}&\hl{{\ref{fpplot}a}}&$\mathhl{1.1\,\times\,10^{-2} {\displaystyle\left(\frac{P_\mathrm{f}}{N_\mathrm{f}} \right)^{\displaystyle 1.0^*}}}$&\hl{0.76} \\
        \hline
        &&&&&\\[-1em]
        ii)&\hl{Gravity currents}&\hl{Dilute ($P_\mathrm{f}$)}&\hl{{\ref{fpplot}b}}&$\mathhl{5.6\,\times\,10^{-3} {\displaystyle\left(\frac{P_\mathrm{f}}{N_\mathrm{f}} \right)^{\displaystyle 0.36}}}$&\hl{0.65} \\
        &&&&&\\[-1em]
        iii)&&\hl{Dilute ($P_\mathrm{th}$, linear fit)}&\hl{-}&$\mathhl{1.0\,\times\,10^{-1} {\displaystyle\left(\frac{P_\mathrm{th}}{N_\mathrm{th}} \right)^{\displaystyle 1.0^*}}}$&\hl{0.34} \\
        &&&&&\\[-1em]
        iv)&&\hl{Dilute ($P_\mathrm{th}$)}&\hl{{\ref{fpplot}c}}&$\mathhl{1.1\,\times\,10^{-2} {\displaystyle\left(\frac{P_\mathrm{th}}{N_\mathrm{th}} \right)^{\displaystyle 0.49}}}$&\hl{0.72} \\
        \hline
        \multicolumn{6}{l}{*The power of correlation is fixed as unity for the linear fits.}
    \end{tabular}
    \caption{Fitted power-law correlation \hl{results} and the coefficient of determination, $R^2$, of the subsets of data points (Fig. \ref{fpplot}), using orthogonal distance regression (see Methods).}
    \label{tab:fitparams}
\end{table}
From the \hl{fluvial flow} data (Fig. \ref{fpplot}a), three regimes of concentration are identified: dilute, transitional, and hyperconcentrated flow.
Dilute flows, \hl{$\Phi \lesssim 10^{-2} $}, are characterised by a \hl{linear} increase in concentration with dimensionless flow power\hl{, $P_\mathrm{f}/N_\mathrm{f}$ (Fig. {\ref{fpplot}}a and Table \ref{tab:fitparams} i).}
\hl{With increasing concentration,  $10^{-2}\lesssim \Phi \lesssim 10^{-1}$, transitional flows exhibit a change in correlation from increasing to decreasing dimensionless flow power as concentration increases.
This transition may be} explained by the onset of turbulence dampening and particle-particle interactions dominating the sediment transport mechanics \cite{vanMaren2009}.
Hyperconcentrated flows, \hl{$\Phi \gtrsim 10^{-1}$}, are \hl{characterised} by a non-linear decrease in dimensionless flow power with increasing concentration\hl{.}
As concentration continues to rise, dimensionless flow power decreases by at least two orders of magnitude.
\par
The \hl{gravity} current data (Fig. \ref{fpplot}b--\hl{e}) \hl{also} exhibit three similar regimes of concentration; the threshold concentrations between regimes are \hl{approximately equal to those of fluvial systems.}
However, the correlation of concentration with dimensionless flow power is remarkably different.
\hl{The dimensionless flow power, based on both the log-law energy production model $P_\mathrm{f}$ (Fig. {\ref{fpplot}}b), and the top-hat model, $P_\mathrm{th}$ (Fig. {\ref{fpplot}}c),} have a strongly non-linear correlation \hl{with sediment concentration, $\Phi$.} 
\hl{The fit of both models is good, but shows an improvement when using the top-hat-based correlation, see Table {\ref{tab:fitparams}} ii) and iv).}
\hl{However, a linear model provides a poor best fit in comparison, contrast Table {\ref{tab:fitparams}} iii) and iv).}
\hl{Moreover, the turbidity current data (Fig. {\ref{fpplot}d}) show almost identical non-linear dependency to the pyroclastic density current data (Fig. {\ref{fpplot}e}), suggesting that the non-linear dependency is universal to all types of gravity currents.}
\par
\hl{Critically,} the non-linear relationship results in dilute \hl{gravity} currents being able to maintain a higher suspended sediment concentration \hl{versus fluvial} flows of an equivalent dimensionless flow power (Fig. {\ref{fpplot}}\hl{c}).
\hl{Previously unrecognised, this has the potential to explain autosuspension in long-runout systems.}
\hl{The correlation suggests that} when a dilute gravity current accelerates, it is not as erosive as a fluvial system\hl{, and similarly, when a gravity current decelerates, deposition is more limited}. 
This implies that \hl{the} suspended-load \hl{of gravity} currents \hl{is} significantly underestimated, \hl{i.e. providing more motive force on shallower slopes}, and \hl{is} less sensitive to changes in flow power than has previously been assumed based on the use of fluvial analogues\cite{Parker1982,Parker1986,Pratson2000debris,KosticParker2006,Hu2009,PantinFranklin2009,Dorrell2018equilibrium,AmyDorrell2021Equilibrium}.
\hl{Since the limited super-dilute pyroclastic density current dataset also exhibits a similar non-linear trend to turbidity currents (Fig. {\ref{fpplot}e}), it is likely that the fluvial-based or top-hat gravity current models are a poor approximation not only for turbidity currents but also for particle-driven gravity currents in general.}
\par
It is noted that a limited proportion of the data (see Supplementary \hl{Table 1}) use cohesive material (kaolinite). 
\hl{It is plausible that flocculation of cohesive particles may occur, increasing particle settling velocity, and decreasing dimensionless flow power.}
However, the growth of flocs is limited by shear rates and their size decreases as the flow increases\cite{Javis2005flocreview}.
Therefore, in strongly sheared \hl{gravity} current experiments, the development of flocs and resulting underestimation of settling velocity is \hl{expected} to be limited.
Further, it would not change the observation that the relationship between concentration and dimensionless flow power is non-linear\hl{, see Fig. \ref{fpplot}b--e, Table \ref{tab:fitparams} ii) and iv)}.
\hl{Moreover, whilst the increase in $P_\mathrm{th}/N_\mathrm{th}$ of gravity currents (Fig. {\ref{fpplot}}c) follows the trend of  fluvial data in the transitional regime it is based on an empirical water entertainment function, $e_\mathrm{w}$.
The empirical water entrainment function has been developed for dilute currents, thus it is expected the values of $P_\mathrm{th}/N_\mathrm{th}$ in the transitional regime have some inherent error.}
\hl{However, this does not impact the primary findings of the non-linear correlation in the dilute regime.}
\section*{Discussion}
\begin{figure}[tp]
    \centering\boxhl{\includegraphics{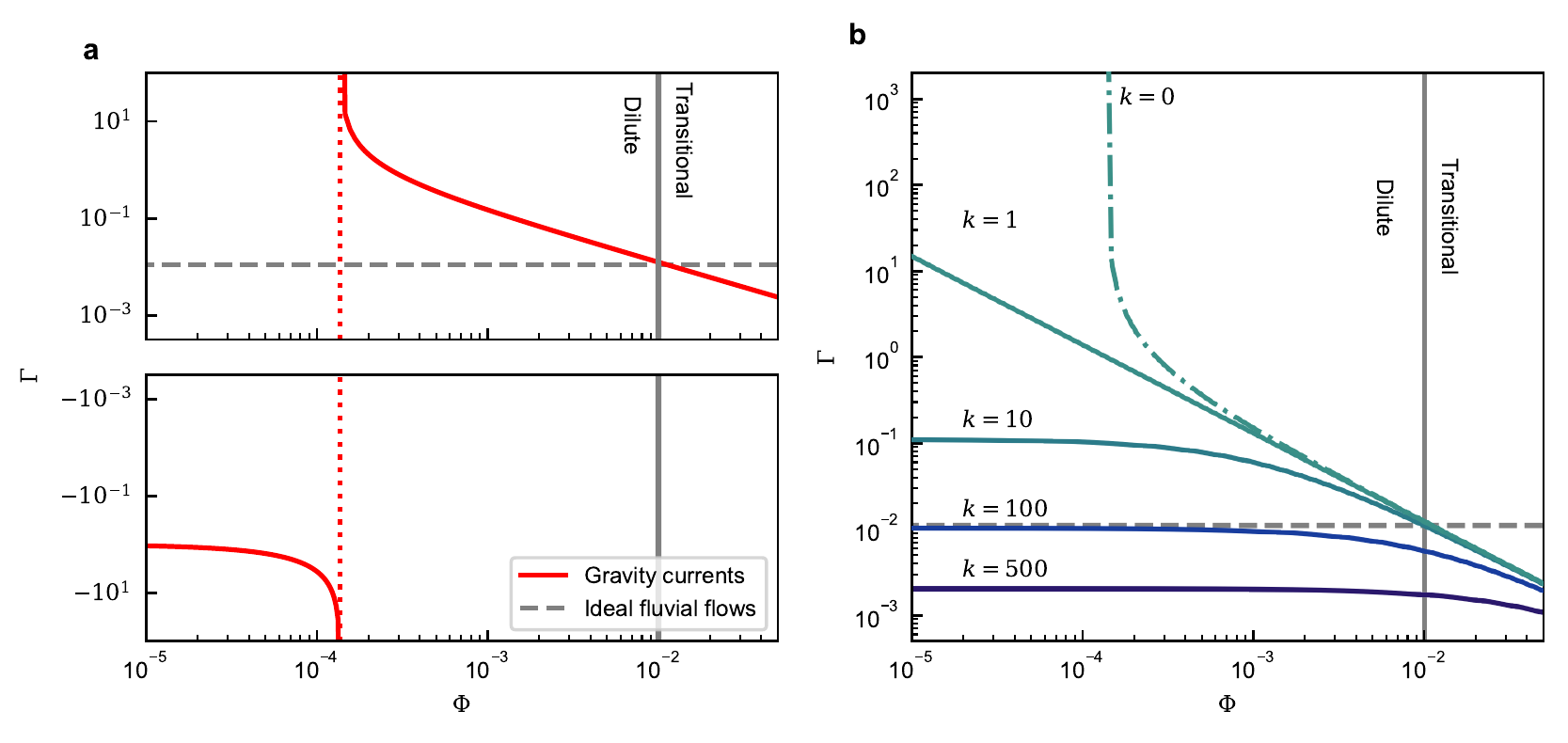}}
    \caption{The turbulent flux coefficient, $\Gamma$ as a function of $\Phi$. 
    a) \hl{Computed $\Gamma$, assuming $P=P_\mathrm{th}$, where it is assumed $S=0$.
    The red dotted line indicates the vertical asymptote for the gravity current curve (red solid).
    } 
    b) \hl{Computed $\Gamma$, assuming $P=P_\mathrm{th}+S$, where the additional energy is parameterised in terms of the buoyancy production $S=kB_\mathrm{th}$}. 
    The dotted-dashed curve \hl{has} $k < 1$, for which $\Gamma$ diverges to infinity for some values of $\Phi$. \hl{Throughout, the Dilute/Transitional regime threshold and the ideal fluvial flow curve is depicted by gray solid and gray dashed lines respectively.}} 
    \label{fig:Gamma}
\end{figure}
The fundamental differences between \hl{fluvial flows and dilute gravity currents (turbidity currents and pyroclastic density currents)} documented above raise the following questions:
\hl{Do the top-hat mean-flow energy loss, buoyancy production and dissipation balance?} What are the implications for autosuspension \hl{models}?
\par
\hl{In top-hat gravity current models\cite{Parker1986}, $P_\mathrm{th}$ is the energy released by the mean-flow and available for the suspension of particles, which requires energy $B_\mathrm{th}$. Additionally, the flow experiences viscous dissipation of energy $\epsilon$. However, as will be shown, there is insufficient energy in this model to suspend the sediment. For the present analysis, the missing energy will be denoted by $S$, so that the energy balance is}
\begin{align}
    \mathhl{P = P_\mathrm{th}} + S \simeq \mathhl{B_\mathrm{th} + \epsilon_\mathrm{th}}  = \left(1+\frac{1}{\Gamma}\right) \mathhl{B_\mathrm{th}},
    \label{eq:PS}
\end{align}
\hl{where the total turbulent flux coefficient, $\Gamma = B_\mathrm{th}/\epsilon_\mathrm{th}$, denotes the ratio of top-hat buoyancy production, $B_\mathrm{th}$, to dissipation, $\epsilon_\mathrm{th}$.}
Using Equation \eqref{eq:PS} and \hl{the curve fitting results, see Table {\ref{tab:fitparams}} i) and iv),} $\Gamma$ can be expressed as:
\begin{align}
    \mathhl{\Gamma = \left(\frac{P_\sim}{B_\sim}+\frac{S}{B_\sim}-1 \right)^{-1}
    \simeq \begin{cases}
        1.1\times10^{-2} &\text{: Fluvial flows}\ (S = 0,\quad B_\sim=B_\mathrm{f},\quad P_\sim=P_\mathrm{f})\\
        \left(9\times10^{3}\,\Phi^{1.0} + k -1\right)^{-1} &\text{: Gravity currents}\ ( B_\sim=B_\mathrm{th},\quad P_\sim=P_\mathrm{th})
    \end{cases}}
    \label{eq:Gamma}
\end{align}
\hl{where $k = S/B_\mathrm{th}$ denotes the ratio of missing energy to the top-hat buoyancy production, $B_\mathrm{th}$.}
\hl{The fluvial data (Fig. {\ref{fpplot}}a) and} Equation \eqref{eq:Gamma} implies that, for fluvial flows, $\Gamma$ is constant. 
Only \hl{$\sim 1.1$\% of the energy} production is consumed by buoyancy production, while the rest is consumed by dissipation.
\hl{For gravity currents in contrast $\Gamma$ depends on $\Phi$ (Fig. {\ref{fig:Gamma}}a).}
\hl{Assuming that the extra energy source, $S=0$ and thus $k=0$, $\Gamma$ grows with decreasing flow concentration, diverging at $\Phi = 1.4 \times 10^{-4}$,} before becoming negative.
However, \hl{$\epsilon_\sim$} and \hl{$B_\sim$} are strictly positive, thus $\Gamma$ must \hl{also always be} positive.
\par
Clearly, $S = 0$ is a poor approximation.
Equation \eqref{eq:Gamma} implies that the \hl{energy} balance of near-equilibrium \hl{gravity} currents can only be satisfied with a non-zero \hl{energy} source/sink, $S$\hl{.}
To satisfy the minimum requirement, $\Gamma > 0$ for all $\Phi$, the additional energy source term is constrained by $k > 1$.
A hypothesized upper limit for the turbulent flux coefficient\cite{Osborn1980epsilon,Caulfield2021} is $\Gamma \mathhl{\leq} 0.2$\hl{, this is broken for $\Phi < 6 \times 10^{-4}$}. 
To satisfy this limit a value of $k \sim 10$ would be required.
\hl{However, it is unlikely that turbidity currents reach this maximum mixing efficiency, and a larger value of $k$ is likely required.}
Previously, it has been assumed that the amount of TKE consumed by buoyancy production in \hl{gravity} currents is similar to that in \hl{fluvial} flows\cite{Pantin1979autosuspension}.
To satisfy $\Gamma \sim 10^{-2}$, Equation \eqref{eq:Gamma}, then $k \sim 100$ (Fig. \ref{fig:Gamma}b)\hl{.} 
\hl{Thus, K-B type criteria and top-hat gravity current models\cite{Bagnold1962,Parker1982,Parker1986,Pratson2000debris,KosticParker2006,Hu2009,PantinFranklin2009,Dorrell2018equilibrium,AmyDorrell2021Equilibrium} fail to explain the energy balance of gravity currents.}

\par
The \hl{energy} balance of \hl{gravity} currents required for autosuspension cannot be explained without an additional \hl{energy source, i.e. $S>0$ in Equation {\eqref{eq:PS}} and Figure {\ref{fig:Gamma}}.}
\hl{Crucially, if the energetic mechanisms} were the same for \hl{gravity} currents and \hl{fluvial} flows, then \hl{gravity currents would be substantially more dilute}, cf. Figure {\ref{fpplot}}\hl{b--e}.
Therefore, to explain autosuspension\hl{, mechanisms for particle uplift must be present that are absent, or of negligible importance, in fluvial flows.}

\begin{figure}[tp]
    \centering
    \boxhl{\includegraphics{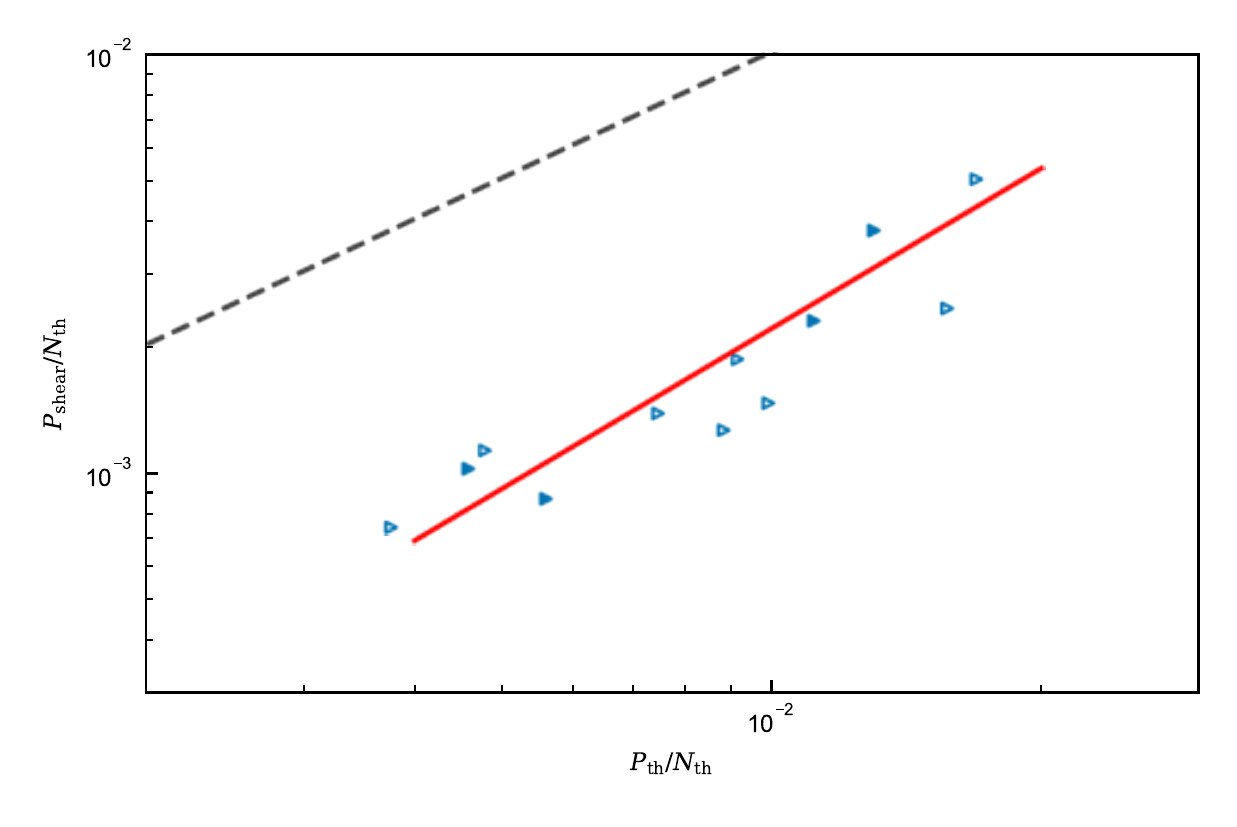}}
    \caption{Log-log plot of depth-integrated \hl{shear} production of TKE against \hl{top-hat total energy loss of a mean flow $P_{\mathrm{th}} = u_*^2U + \frac{1}{2}e_\mathrm{w}U^3$.}
    \hl{The} power-law correlation (Eq. \ref{eq:empiricalPll}) is depicted by a red solid line. 
    Gray dashed line indicates the line of equality \hl{($P_\mathrm{shear} = P_\mathrm{th}$)}. 
    Data are calculated from the empirical measurements reported by Islam and Imran\cite{IslamImran2010}. 
    Symbols as per Figure \ref{Shields}.}
    \label{bed-shear and total-shear}
\end{figure}

\hl{It is plausible that the shear production, Equation \eqref{eq:turbulent_production}, predicted by top-hat models is less than the actual production in real flows, and this possibility is addressed first. The shear production is calculated from empirical data\cite{IslamImran2010}, and plotted in Figure \ref{bed-shear and total-shear}. However, the limited data available suggest that the directly computed shear production term, $P_\mathrm{shear}$, is substantially lower than the `top-hat' energy loss term, $P_\mathrm{th}$,}
\begin{align}
    \mathhl{\frac{P_\mathrm{shear}}{N_\mathrm{th}}} &= \mathhl{0.74 \left( \frac{P_\mathrm{th}}{N_\mathrm{th}} \right)^{1.27}}
&
    \mathhl{(R^2} &\mathhl{= 0.92)}.
    \label{eq:empiricalPll}
\end{align}
\hl{Consequently, if  shear production represents all available energy, the size of the missing energy source is substantially larger.}

\begin{figure}[tp]
    \centering
    \boxhl{
    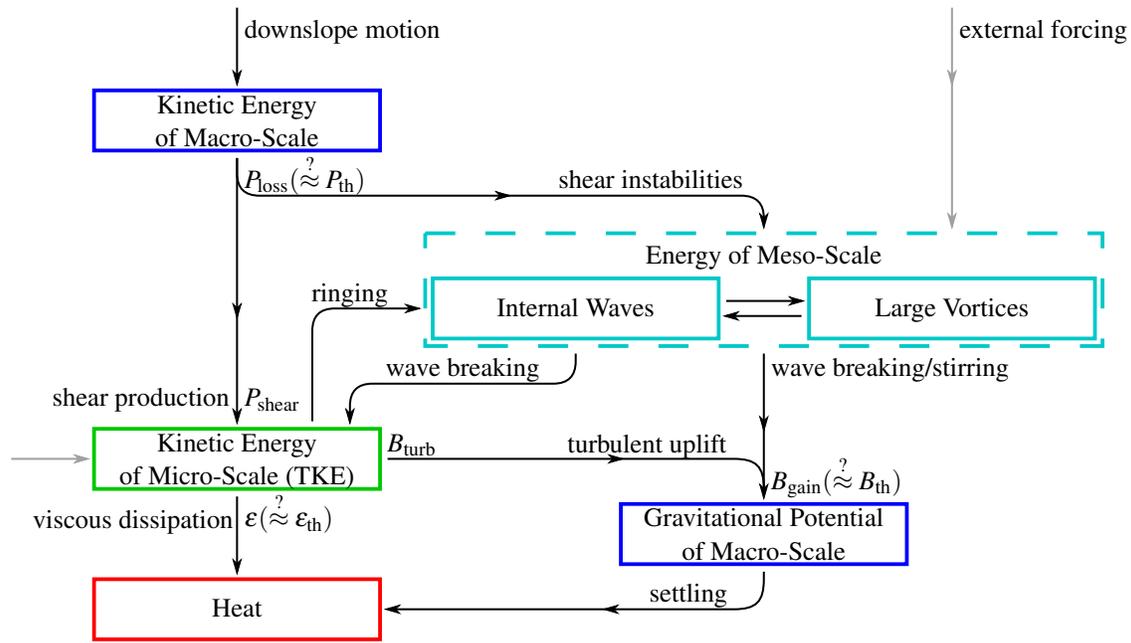
    }
    \caption{\hl{The stores of energy within gravity currents, (blue boxes) and the transfer of energy between them (black arrows). The energy lost at the tail of each arrow is equal to the energy gained at its tip. Grey arrows represent the input of energy from external forcing (energy could also be lost to these forces).}}
    \label{fig:energetics}
\end{figure}

\hl{To understand the origin of these shortcomings, the energetic dynamics within a gravity current will be broken down and explored conceptually; a schematic of these dynamics is provided in Figure \ref{fig:energetics}.
The dynamics of the flow occur on three distinct length-scales: the macro-scale flow on the scale of the length of the current, which is the scale of the top-hat model; the meso-scale flow on the scale of the depth of the current, which is able to support internal waves\cite{lefauve2018holmboe,Kostaschuk2018pulse,Dorrell2019,Marshall2021,Salinas2021} and the largest vortices\cite{Peakall2015,Dorrell2018mixing}; and the micro-scale flow, which supports the turbulent vortices (TKE). In Figure \ref{fig:energetics} the gravitational potential of the sediment is split between the macro-scale contribution, the vertical distribution of the sediment is (on average) slowly varying, and the meso-scale contribution due to the rise and fall of internal waves, for example.
}

\hl{
In simplified models of gravity currents, such as top-hat models, it is the macro-scale kinetic energy that is captured, and on shallow slopes it is this macro-scale flow which is energised by the down-slope component of gravity.
As the longitudinal flow accelerates/decelerates, the flow thins/thickens, exchanging the macro-scale energy between the kinetic and gravitational potential (not included in Figure \ref{fig:energetics}).
This kinetic energy is lost at a rate $P_\mathrm{loss}$.
Large scale internal shear can generate flow instabilities, such as the Kelvin-Helmholtz or Holmboe instabilities, resulting in internal waves{\cite{lefauve2018holmboe,Kostaschuk2018pulse,Dorrell2019,Marshall2021,Salinas2021}}, which can be seeded at flow initiation\cite{Best2005,Kostaschuk2018pulse}. 
Alternatively, the mean flow energy may be used to stimulate large vortices, for example, secondary flow circulation\cite{Straub2011sinuous,Peakall2015,Dorrell2018mixing}.
The internal shear generated by the macro-scale flow also directly energises turbulent vortices through shear production.
The meso-scale flow structures are able to generate gravitational potential directly, by stirring the flow\cite{Straub2011sinuous}, and through wave breaking which also generates TKE. 
The turbulence is, in turn, able to `ring' density interfaces generating internal waves{\cite{Dorrell2019,Marshall2021, lloyd_dorrell_caulfield_2022}}, or uplift particles through the diffusive effect of vortices. 
The TKE and gravitational potential are slowly lost to heat though viscous effects. 
}

\hl{In addition to these internal processes, in many currents external forcing directly drives the flow, also depicted in Figure \ref{fig:energetics}. Examples include the Coriolis force\cite{Dorrell2013,Davarpanah2020Coriolis}, bottom currents\cite{Miramontes2020}, tidal forcing, return flows\cite{Sumner2014helical,Toniolo2007ponded,Patacci2015ponded}, and thermal effects in pyroclastic density currents\cite{Andrew2012PDC,Shimizu2019}. However, these effects are not equally featured in the flows in our dataset, and are expected to result in quite different internal dynamics. Thus, the trend in Figure \ref{fpplot}(c) and the derived missing energy cannot be explained by these external forces, and instead they generate the scatter about the trend.}

\hl{It is worth pausing to reconsider what the top-hat model represents. The production, $P_\mathrm{th}$, is derived as approximating the energy loss from the mean flow $P_\mathrm{loss}$, and $B_\mathrm{th}$ as approximating the energy gain by the sediment $B_\mathrm{gain}$. However, it is clear from Figure \ref{fpplot} and the implied missing energy that at least one of these approximations is inaccurate, most likely both. Through Figure \ref{bed-shear and total-shear} we see that $P_\mathrm{th}$ does not approximate the shear production. Consequently, the top-hat production does not represent \emph{any} of the indicated production terms on Figure \ref{fig:energetics}.}

\hl{Therefore, a likely explanation (indeed the only one that remains) is that the top-hat model is wrong: the dynamics of the flow are not well represented if the information about the vertical variation of density and velocity are neglected.
The origins of the top-hat model are in fluvial systems, where the log-law of the wall holds across a large portion of the flow. 
In this context, it is possible to calculate the shear production $P_\mathrm{shear}$ exactly\cite{Bagnold1962,Parker1986,Pope2006,Andersen2007,AmyDorrell2021Equilibrium}, and the result is here denoted by $P_\mathrm{f} = u_\star^2 U$. 
A consequence of the driving density being uniform over the depth of the flow is that the top-hat model, in fluvial systems, gives exactly the same energy loss from the kinetic energy of the mean flow\cite{Rastogi_Rodi1978OC}, which enables the top-hat models to accurately capture the flow energetics (up to internal waves and large-scale vortices). 
There is no reason to believe that the energetics of gravity currents are similarly well represented by top-hat models because the driving density varies over the depth.
Experiments have shown that the shape factors resulting from the depth variation of density and velocity in laboratory currents differ from top-hat models by up to $40 \%$\cite{IslamImran2010}.
In other settings, a strong density interface is generated around the maximum velocity, which is maintained through locally negative turbulent production driven by radiation stresses\cite{Dorrell2019,Marshall2021,Salinas2021} analogous to atmospheric jets\cite{Dritschel2011}.
In either case, the loss of energy from the mean flow will be substantially different to the top-hat model. It is possible that $P_\mathrm{loss}>P_\mathrm{th}$, a large portion of this energy loss would need to go into the meso-scale structures because the mean-flow energy loss would be substantially larger than the shear production. 
In addition, the buoyancy production required by top-hat models is an upper bound, the sediment is assumed to be as high up as possible, and including the vertical variation of density would reduce this requirement so that $B_\mathrm{gain}<B_\mathrm{th}$. 
The reduction in the expected amount of energy passing through the TKE budget reduces the expected amount of energy loss to dissipation, meaning that a larger portion of the energy loss by the mean flow goes into particle uplift.}

\hl{While investigations into the vertical structure of gravity currents have been conducted\cite{Parker1987,IslamImran2010}, the implications for the energetics remains an open problem. The momentum balance and sediment transport models may also need to be updated. Here, the vertical structure has already been incorporated in models,\cite{Parker1987}, though it is not clear if this is sufficient to capture the effective force generated during the production of meso-scale structures, and the resulting dissipation-free uplift of particles.}

\hl{The explanations can be summarised by writing}
\begin{align}
    \mathhl{S = (P_\mathrm{loss}-P_\mathrm{th}) + (B_\mathrm{th}-B_\mathrm{gain}) + (\epsilon_\mathrm{th}-\epsilon),}
\end{align}
\hl{which summarises the effective extra energy avaliable in a real current compared to the modelled current. Here, $P_\mathrm{loss}>P_\mathrm{th}$ due to the additional macro-scale kinetic energy lost to meso-scale internal waves and large vortices. 
This invalidates the long-stannding Knapp-Bagnold hypothesis that all the energy lost by the macro-scale flow drives turbulence through shear production, and that turbulence is the only means of particle uplift. 
Note that the additional energy loss does not necessarily imply an energy depletion, real gravity currents may have a larger macro-scale kinetic energy budget than top-hat models due to the vertical variation of velocity.
Additionally, $B_\mathrm{gain}< B_\mathrm{th}$, due to the lowering of the centre of mass when the vertical structure of density is captured\cite{Dorrell2014}, which reduces the gravitational potential that must be maintained. Finally, $\epsilon<\epsilon_\mathrm{th}$ because the TKE budget is lower, a large portion of the energy instead stored in the meso-scale structures. Thus, the presence of meso-scale structures increases the energetic efficiency of autosuspension.}

\hl{
To address the apparent missing energy in the models, future work must move beyond the approximations appropriate for open-channel and fluvial systems, and capture the complexity present in the structure and internal dynamics of gravity currents. The resulting additional capacity to support particles, that is the increased autosuspension capability, has numerous implications for environmental currents. The long run-out of turbidity currents has been a long standing enigma, and the results presented here show that the current is able to maintain a much higher sediment load than previously believed. This gives significantly more driving force on shallow slopes, and a much slower deposition rate of particles, which facilitates the transport of sediment to the distal parts of submarine systems. More broadly, particle-driven gravity currents are known to be highly destructive, with flows capable of causing immense damage. For the accurate prediction of gravity currents, this work shows that research focus is required on the dynamics of meso-scale energy exchange and balances, to be captured by the next generation of reduced order models.
}

\section*{Methods}
\begin{figure}[tp]
\centering
\includegraphics{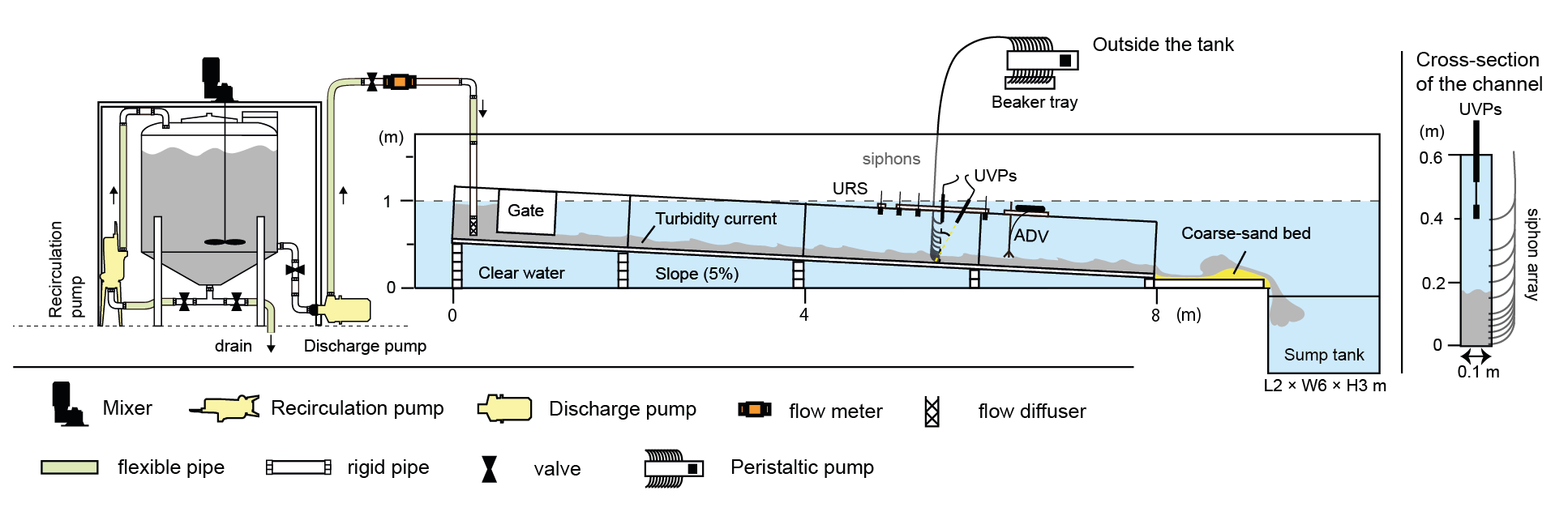}
\caption{Schematic illustration of the flume setup for the experiments. \hl{The f}low sampling location is at 4.7 m from the inlet (5.7 m downstream from the upstream end of the channel). The cross-section of the channel at the measurement location is depicted on the right side of the figure.}
\label{exp}
\end{figure}
To evaluate the controls on the transport of sediment by turbidity currents, the dynamics of pseudo-steady state, and turbulent flows are examined.
Both laboratory and real-world data are combined to cover a range of scales.
To constrain the dynamics to pseudo-steady flows, only empirical data of continuous discharge or long-duration flows are considered.
Rapidly varying flows, such as lock-exchange experiments or short-duration field observations, are omitted. 
\begin{table}[htbp]
\begin{center}
\begin{tabular}{c c c c c c c c c c} 
 \hline
 Experiment & $d_{50}$ ($\rm \mu m$) & $\phi_0$ (vol.\%) & $Q$ (l/s) & $D_{\rm flow}$ (s) & $T_{\rm ambient}$  & $T_{\rm flow}$ ($^{\circ}$C) & $\log_{10}\mathrm{Re}$ & $\mathrm{Ri}$ & $\mathrm{Re_p}$ \\ [0.5ex] 
 \hline
 01 & 40 & 15.4 & 6 & 160 & 14.0 & 14.5 & 4.8 & 0.92 & 1.03 \\
 
 02 & 42 & 13.3 & 6 & 160 & 14.2 & 15.2 & 4.8 & 0.98 & 1.11  \\
 
 03 & 42 & 11.7 & 6 & 160 & 13.5 & 14.3 & 4.8 & 0.73 & 1.1 \\
 
 04 & 41 & 10.0 & 6 & 160 & 13.3 & 14.5 & 4.9 & 0.73 & 1.04\\
 
 05 & 43 & 8.0 & 6 & 160 & 13.3 & 15.0 & 4.9 & 0.84 & 1.14\\
 
 06 & 41 & 5.7 & 6 & 160 & 13.3 & 14.1 & 4.8 & 1.07 & 1.07 \\
 
 07 & 43 & 4.5 & 4 & 240 & 15.0 & 15.4 & 4.8 & 1.16 & 1.14\\
 
 08 & 41 & 1.7& 4 & 240 & - & - & 4.8 & 1.16 & 1.07\\
 
 09 & 43 & 1.2& 4 & 240 & - & - & 4.9 & 0.93 & 1.14\\
 
 10 & 43 & 0.3& 4 & 240 & 14.2 & 14.7 & 4.8 & 0.78 & 1.14\\
 
 11 & 38 & 1.6 & 4 & 240 & 13.9 & 14.4 & 4.8 & 1.05 & 0.94 \\
 
 12 & 33 & 0.5& 4 & 240 & 13.9 & 14.5  & 4.8 & 1.5 & 0.77\\ 
 
 13 & 9 & 4.0 & 3.5 & 280  & 16.2 & 16.7 & 4.6 & 0.68 & 0.11\\

 14 & 9 & 6.3 & 3.5 & 280 & 16.4 & 17.7 & 4.7 & 0.56 & 0.11 \\ 
 
 15 & 9 & 8.0 & 3.5 & 280 & 17.6 & 15.1 & 4.7 & 0.56 & 0.11\\ 
 
 16 & 9 & 9.6 & 3.5 & 280 & 17.6 & 16.4 & 4.7 & 0.63 & 0.11\\
 
 17 & 9 & 12.2 & 3.5 & 280 & 15.9 & 17.9 & 4.8 & 0.68 & 0.11\\[1ex] 
 \hline
\end{tabular}
\caption{Experimental conditions. Median particle size  ($d_{50}$), initial concentration in the mixing tank ($\phi_0$), discharge rate ($Q$), total flow duration ($D_{\rm flow}$), the temperature of ambient water ($T_{\rm ambient}$) and of the flow ($T_{\rm flow}$), Reynolds number ($\mathrm{Re}$), Richardson number ($\mathrm{Ri}$), and particle Reynolds number ($\mathrm{Re_p}$) are listed. $T_{\rm ambient}$ and $T_{\rm flow}$ are calculated by the acoustic doppler velocimeter (ADV) as time-averaged values. The turbidity currents for the experimental runs are characterized by either glass beads (01--12) or kaolinite (13--17).}
\label{table:Exp}
\end{center}
\end{table}

\paragraph{Data analysis} 
Numerous studies report detailed vertical profiles of flow velocity and concentration data for gravity flows, including both sediment-laden turbidity currents and conservative composition-driven flows. 
The data for composition-driven flows are not used in the main analysis in this study, but instead to validate the applicability of the developed profile interpolation/extrapolation methodology (see \hl{Supplementary Figure 3--4}).
Furthermore, three different datasets from direct measurements\cite{Simmons2020} of flows in the Congo canyon are added to this study to compare laboratory-scale and natural-scale turbidity currents.
Since each study considered here uses different materials and methodologies, the comparison between datasets was based on consistent data interpolation and extrapolation methods to reconstruct the full vertical profiles of the flow (see \hl{Supplementary Note 2}). 
In this study, the compiled data were categorized into two types: TYPE I) the gravity current data in which both vertical velocity and concentration profiles are measured, and TYPE \hl{II}) gravity current data in which velocity profiles are available but vertical concentration profiles are not available.
\par
Approximately two-thirds of compiled sources and the flume experiments in this study provide vertical profiles of both streamwise velocity and flow concentration (TYPE I: See \hl{Supplementary Table 1}). For the remaining sources, either a part of or all of the reported experiments do not provide vertical concentration profiles (TYPE \hl{II}). 
The depth-averaged flow concentration is estimated for those sources as follows. 
First, using the data that included both the concentration in the mixing tank and that measured in the flume (TYPE I), we construct an empirical relationship between these two quantities (see Supplementary Figure 9). 
Then, the concentration in the flume (for those experiments that did not report it: TYPE II) is assumed to exactly satisfy the constructed relationship.
\par
From the interpolated and extrapolated profiles, flow parameters are computed, including depth-averaged velocity, concentration, and flow depth (see \hl{Supplementary Note 1}).
The median particle size, $d_{50}$, is used to calculate the settling velocity of each experiment.
For sand-sized particles, an empirical formula covering a combined viscous plus bluff-body drag law for natural irregular sand particles\cite{Soulsby1997} is used, and for finer particles, Stokes' law is applied (\hl{see Supplementary Equation (8)}). 
For those data which used the sieving
combined with hydrometer method (SHM) for the calculation of particle-size distribution, the reported median particle $d_{50}$ is amended based on the empirical relation between SHM and the laser diffraction method to avoid the overestimation of clay fraction (see \hl{Supplementary Figure 1} in online supplementary material).

\paragraph{Flume experiments.}
The compiled dataset covers a wide range of flow concentrations ($10^{-2}$--$10^{1}$ vol.$\%$), yet there are two data gaps, one of which is around 0.3--0.6 vol.$\%$ and the other is around 2--4 vol.$\%$ (see \hl{Supplementary Figure 11}).
These gaps motivated new flume experiments, to better investigate the flow power balance of equilibrium turbidity currents.
The experiments were conducted to study the sediment-load capacity of turbidity currents in an idealised channel \cite{decala2020} in the Total Environment Simulator, at the University of Hull.
The main channel is 8 m long, 0.1 m wide, 0.6 m deep with a 5\% slope, and it is submerged in a large water tank 12 m long, 6 m wide, 1 m deep filled with ambient water. 
At downstream, the channel is connected smoothly to a region of coarse sand of area 3 m $\times$ 3 m and 5 cm deep.
A sump tank (2 m long $\times$ 6 m $\times$  3 m) is located at the downstream end of the large tank, to minimize backwater effects\cite{decala2020}.
\par
Sediment-water mixtures are fed into the flume from a 1.0 $\rm m^3$ mixing tank (Fig. \ref{exp}). 
Then, each experimental flow is discharged from the flow diffuser pointed upstream to create a sediment-water cloud which generates a turbidity current by its negative buoyancy.
The initial conditions for each run are set to fill the aforementioned two data gaps (Table \ref{table:Exp}).
Velocity and density profiles are measured 4.7 m downstream from the inlet (Fig. \ref{exp}). 
Velocity measurements are made using two \hl{Met-Flow} Ultrasonic Velocity Profilers (UVPs), mounted at different angles: bed-normal and 30 degrees to the bed-normal angle.
Suspended sediment samples are collected using a multi-channel peristaltic pump (Watson Marlow) connected to a 12-siphon array at a constant sampling rate (21 ml s$^{-1}$).
Siphons are connected to a series of holes on the sidewall of the channel (0.7, 2.8, 4.7, 6.8, 9.0, 11.0, 16.0, 21.0, 26.0, 30.5, 35.7, and 40.5 cm above the bed, see Fig. \ref{exp}) to minimize flow obstruction. 
To measure the aggradation rate, ultrasonic sensors and GoPro cameras are used near the measurement location (see \hl{Supplementary Note 3} for details).
\section*{Acknowledgement}
SF, MGWV, EWGS, EB, WDM and RMD were supported by the Turbidites Research Group, University of Leeds funded by AkerBP, CNOOC group, ConocoPhillips, Murphy Oil, OMV, Occidental Petroleum. RMD was also supported by the UK Natural Environment Research Council [NE/S014535/1].
DRP, EB, RF, and XW were supported by the European Research Council under the European Union's Horizon 2020 research and innovation program [grant 725955]. RF was also supported by the Leverhulme Trust, Leverhulme Early Career Researcher Fellowship [grant ECF-2020-679]. 
We thank Brendan J. Murphy, Fiona Chong, Anne Baar, and Steve M. Simmons for their assistance with the flume experiments at the Deep.
We thank Steve M. Simmons for sharing the observation data from the Congo canyon.
\hl{We are grateful for the UVP-DUO-MX equipment used herein, which was provided in collaboration with Met-Flow.}

\section*{Author contributions}
SF undertook the data compilation and the statistical analyses.
EWGS, EB, HN, DRP and RMD contributed to interpretation of the results of the data compilation and the statistical analyses.
EWGS developed the curve fitting methodology and conducted the curve fitting.
SF, MGWV, EB, RF, XW and RMD designed and conducted the flume experiments on which this study is based on.
All authors contributed to drafting and editing the manuscript.

\section*{Data Availability} Supplementary Information accompanies this paper
\section*{Code Availability} \hl{The algorithms for curve fitting, interpolation/extrapolation of flow profiles, and formulas for calculating each flow parameter are fully described in the Supplementary Information (Note 1,2 and 5).}
\section*{Competing interests} The authors declare no competing interests

\bibliography{Reference_main.bib}

\end{document}


\maketitle
\flushbottom
\thispagestyle{empty}

\section{Non-dimensional flow characterization}
Here the quasi-equilibrium flow condition is assumed, where the mean flow parameters do not vary in time and space but the turbulent fluctuation from mean flow parameters are still present. 
To describe these flow states, coordinate systems, and flow characteristics are introduced in this section.
$\mathbf{x} = (x,y,z)$ denotes the spatial coordinate of the given point, where $x$ is the streamwise direction, $y$ is the lateral direction, and $z$ is the bed-normal direction ($z=0$ at the bed). 
Flow velocity is denoted by $\mathbf{u}(\mathbf{x},t) = \left(u,v,w \right) $. 
Volumetric flow concentration at a given height is denoted by $\phi(z)$. Reynolds averages\cite{pope2000turbulent} are introduced here to describe the mean flow characteristics and their fluctuations as follows
\begin{align}
    u(\mathbf{x},t) &= \langle u(\mathbf{x},t)\rangle + u^\prime(\mathbf{x},t),\ \
    v(\mathbf{x},t) = \langle v(\mathbf{x},t)\rangle + v^\prime(\mathbf{x},t),\\
    w(\mathbf{x},t) &= \langle w(\mathbf{x},t)\rangle + w^\prime(\mathbf{x},t),\ \
    \phi(\mathbf{x},t) = \langle \phi(\mathbf{x},t)\rangle + \phi^\prime(\mathbf{x},t).
    \label{eq:reynolds-equations}
\end{align}
where $\langle\cdot\rangle$ denotes the temporal Reynolds average and primes denote the Reynolds fluctuations ($\langle u^\prime \rangle = \langle v^\prime \rangle = \langle w^\prime \rangle = \langle \phi^\prime \rangle = 0 $). 
Then, the mean-flow field of quasi-equilibrium turbidity current is defined by introducing depth-averaged parameters as follows
\begin{align}
    U = \frac{1}{h(x)}\int_0^\infty \langle u(\mathbf{x})\rangle {\rm d}z,\ \
    V = \frac{1}{h(x)}\int_0^\infty \langle v(\mathbf{x})\rangle {\rm d}z,\ \
    W = \frac{1}{h(x)}\int_0^\infty \langle w(\mathbf{x})\rangle {\rm d}z,\ \
    \Phi = \frac{1}{h(x)}\int_0^\infty \langle \phi(\mathbf{x})\rangle {\rm d}z,\ \
    \label{eq:depth-averaged-equations}
\end{align}
where $h(x)$ denotes flow depth. 
It should be noted that it is assumed that flow parameters do not vary in time for quasi-equilibrium turbidity currents. 
Flow depth is defined as the height at which both flow velocity and concentration vanish, which is
\begin{align}
    h(x) =  \max \left[ z|_{\langle u\rangle = 0} ,z|_{\langle \phi\rangle = 0}\right],
    \label{eq:flow-depth}
\end{align}
where $z|_{\langle u\rangle = 0}$ is the flow depth above which flow velocity becomes negligible and $z|_{\langle \phi\rangle = 0}$ is the flow depth above which concentration difference between the floa and the ambient fluid becomes negligible.
In the ideal condition, $z|_{\langle u\rangle = 0} = z|_{\langle \phi\rangle = 0}$.
However, in the actual measurement of experiments, $z|_{\langle u\rangle = 0}$ is not always the same value as $z|_{\langle \phi \rangle = 0}$ for various reasons such as i) the effects of the non-zero flow velocity of ambient fluid, ii) the error of velocity or density measurements at the upper layer of the flow, and iii) error due to the extrapolation methodology. 
In particular, \hl{for turbidity currents}, the vertical gradient of concentration \hl{profiles}  tends to be very small and show gradual transition to \hl{the} ambient fluid. 
Thus, measurement error near the upper boundary could cause a significant error of the estimation of $z|_{\langle \phi \rangle = 0}$. 
The majority of flume experiments use a few to a dozen siphon tubes to measure the flow concentration. 
On the other hand, the majority of compiled experiments measured vertical velocity profiles by using UVP or ADV, which have a much higher resolution than a siphon array.
However, some of the flume experiments also have low resolution of velocity profiles, such as Michon et al.\cite{Michon1955exp}, Altinakar et al.,\cite{Altinakar1988thesis} and Tesaker\cite{Tesaker1969uniform} (Table \ref{table:source}). 
Thus, the accuracy of $z|_{\langle u \rangle = 0}$ and $z|_{\langle \phi \rangle = 0}$ varies between each experiment. 
In this study, the zero-velocity and zero-concentration heights are carefully compared, and the more reliable one was employed as flow depth (see below for detailed procedures).
\par
To describe the flow state, the following non-dimensional flow parameters are introduced. Firstly, the Richardson number, $Ri$ the ratio of the buoyancy and the flow shear stress, is given as,
\begin{align}
    \mathrm{Ri} = \frac{-g (\partial \mathhl{\langle \rho\rangle}/\partial z)}{\mathhl{\rho_\mathrm{a}} (\partial\langle u \rangle/\partial z )^2 } \simeq \frac{g\Delta \rho/h}{\mathhl{\rho_\mathrm{a}} U^2/h^2} = \frac{gR\Phi h}{U^2}
    \label{eq:Richardson-number}
\end{align}
where $\rho$ is the flow density, \hl{$\rho_\mathrm{a}$} is the density of the ambient fluid, and $\Delta \rho$ is the density difference between the flow and the ambient fluid \hl{($\langle\rho\rangle - \rho_\mathrm{a}$)}. 
The shear velocity, $u_*$, is calculated using the logarithmic law local to the bed,
\begin{align}
    \frac{{\rm d}\langle u \rangle}{{\rm d} z} = \frac{u_*}{\kappa z}
    \label{eq:ustarfromgradient}
\end{align}
The skin drag coefficient, $C_\mathrm{D}$, is defined as
\begin{align}
    C_\mathrm{D} = \frac{u_*^2}{U^2}.
    \label{eq:skindrag}
\end{align}
In this study, the particle settling velocity, $w_\mathrm{s}$, for relatively coarse material (> $100 {\rm \mu}$m) is prescribed by an empirical formula covering a combined viscous plus bluff-body drag law for natural irregular sand grains \cite{Soulsby1997},
\begin{align}
    w_\mathrm{s} = \frac{\nu}{d_{50}}\left[ (10.36^2 + 1.049D_\mathrm{s}^3)^\frac{1}{2} - 10.36 \right].
    \label{eq:settling-velocity}
\end{align}
where $d_{50}$ is the median grain size, and $D_\mathrm{s} = (g\Delta \rho /\rho \nu^2)^{1/3}d_{50}$ is the dimensionless particle diameter, $\nu$ is the fluid viscosity. For the finer particles (< $100 {\rm \mu}$m), Stokes' law is applied,
\begin{align}
    w_\mathrm{s} = \frac{Rgd_{50}^2}{C_1\nu}
    \label{eq:settling-velocity_stokes}
\end{align}
where $R$ is specific gravity and $C_1$ is a constant with a theoretical value of 18.

\section{Data compilation}

\begin{longtable}{ l | l | c | l | c | l } 
 \hline
 source & TYPE & Slope (\%) & Material & $ d_{50}\ ( {\rm \mu m})$ & Measurement tools \\ [0.5ex] 
 \hline 
 Michon et al.\cite{Michon1955exp} & \hl{I} & 0.3--3.6 & Kaolinite & 14.6 & $\langle u \rangle:$ MPCM\\
  & & & & & $\langle \phi \rangle:$ Siphon array (4--12) \\[0.5ex]
  
 Tesaker\cite{Tesaker1969uniform} & \hl{I} & 5.0--12.5 & Quartz & 360--410 & $\langle u \rangle:$ Velocity meters$^{*6}$(3) \\
  &  &  & Kaolinite & 1.1--1.5$^{*4}$ & $\langle \phi \rangle:$ Siphon array (3)  \\[0.5ex]
  
 Altinakar\cite{Altinakar1988thesis} & \hl{I} & 1.0--2.96 & Quartz & 14 / 32 & $\langle u \rangle:$ MPCM\\ 
   &   &   & Salt  & --  & $\langle \phi \rangle:$ Siphon array (16) \\[0.5ex]
   
 Garc\`ia\cite{Garcia1993} & \hl{I} & 8 & Silica & 9 & $\langle u \rangle:$ MPCM\\
   &   &   &  &  & $\langle \phi \rangle:$ Optical probes\\[0.5ex]
   
 Packman \& Jerolmack\cite{Packman2004TC} & \hl{I} & 1.0 & Quartz & 12 & $\langle u \rangle:$ ADV\\
   &   &   & Kaolinite & 1.3 & $\langle \phi \rangle:$ Siphon array (6) \\[0.5ex]
   
 Amy et al.\cite{Amy2005exp} & \hl{I} & 5.2 & Glycerol & -- & $\langle u \rangle:$ UVP array (8) \\
  &  &  &  &  & $\langle \phi \rangle:$ Siphon array (5) \\[0.5ex]
  
 Islam \hl{\&} Imran\cite{IslamImran2010} & \hl{I \& II} & 0.0-8.0 & Silt / Salt & 25 & $\langle u \rangle:$ ADV   \\
   &   &   &   &   & $\langle \phi \rangle:$ Siphon array (20) \\ [0.5ex]
 
 Sequeiros et al.\cite{Sequeiros2010bedload} & \hl{I} & 0.85--5.0 & Salt$^{*3}$ & -- & $\langle u \rangle:$ ADV \\ 
   &  &  &  &  & $\langle \phi \rangle:$ Siphon array (10) \\ [0.5ex]
 
 Sequeiros et al.\cite{Sequeiros2010ExpProfile} & \hl{I} & 5.0 & Salt$^{*3}$ & -- & $\langle u \rangle:$ ADV \\
   &   &    &   &   &  $\langle \phi \rangle:$ Siphon array (10)\\[0.5ex]
 
 Cartigny et al.\cite{Cartigny2013concentration} & \hl{II} & 12.3--21.3 & Quartz & 160 & $\langle u \rangle:$ Angled UVP \\
   &   &   &   &   & $\langle \phi \rangle:$ --  \\[0.5ex]

 \hl{Varjavand et al.{\cite{Varjavand2015}}} & \hl{I} & \hl{1.25} & \hl{Kaolinite} & \hl{13.4} & \hl{$\langle u \rangle:$ Angled UVP} \\
   &   &   &   &   & \hl{$\langle \phi \rangle:$ Siphon array (14)}  \\[0.5ex]

 Fedele et al.\cite{Fedele2016exp} & \hl{I} & 8.7--17.6 & Salt & -- & $\langle u \rangle:$ ADV \\
   &   &   &   &   & $\langle \phi \rangle:$ Siphon array (20) \\[0.5ex]
 \hl{Breard \& Lube{\cite{BreardLube2017}}} & \hl{I (PDCs)} & \hl{15.8} &\hl{Ignimbrite} & \hl{250} & \hl{$\langle u \rangle:$ PIV} \\
   &   &   &   &   & \hl{$\langle \phi \rangle:$ LC \& PT }\\[0.5ex]
 Leeuw et al.\cite{Leeuw2018exp} & \hl{I} \& \hl{II} & 15.8--19.4 & Sand & 141 & $\langle u \rangle:$ Angled UVP \\
  &  &  &  &  & $\langle \phi \rangle:$ Siphon array (4) \\ [0.5ex]
 
 Leeuw et al.\cite{Leeuw2018channelExp} & \hl{II} & 19.4 & Sand & 131 & $\langle u \rangle:$ Angled UVP   \\ 
  &   &   &   &   & $\langle \phi \rangle:$ --   \\ [0.5ex]
 
 Hermidas et al.\cite{Hermidas2018exp} & \hl{II} & 10.5--16.7 & Quartz & 150/46 & $\langle u \rangle:$ Angled UVP \\
  &   &  & Kaolinite & 0.18 & $\langle \phi \rangle:$ -- \\[0.5ex]

Sequeiros et al.\cite{Sequeiros2018profile} & \hl{I} & 9.0 & Plastic & 57 & $\langle u \rangle:$ Angled UVP \\
 &  &  &  &  & $\langle \phi \rangle:$ Siphon array (11) \\[0.5ex]
 
 Eggenhuisen et al.\cite{Eggenhuisen2019exp} & \hl{I} & 7.0--14.0 & Sand & 130 & $\langle u \rangle:$ Angled UVP  \\
  &   &   &   &   & $\langle \phi \rangle:$ Siphon array (4)  \\[0.5ex]

\hl{Farizan et al.{\cite{Farizan2019}}} & \hl{I} & \hl{1.0} & \hl{Kaolinite} & \hl{11} & \hl{$\langle u \rangle:$ ADV}  \\
  &   &   &   &   & \hl{$\langle \phi \rangle:$ ADV}\\[0.5ex]

 Kelly et al.\cite{Kelly2019} & II$^{*1}$ & 3.5 & Salt & -- & $\langle u \rangle:$ ADV \\
   &  &   &   & -- & $\langle \phi \rangle:$  --  \\[0.5ex]
 
 Koller et al.\cite{Koller2019} & \hl{I$^{*2}$} & 0.9--2.6 & Salt & -- & $\langle u \rangle:$  UVP array (10)\\
   &   &   &   &  & $\langle \phi \rangle:$  Siphon array (6)\\[0.5ex]
   
 \hl{Brosch \& Lube{\cite{BroschLube2020}}} & \hl{I (PDCs)} & \hl{10.5} &\hl{Ignimbrite} & \hl{245} & \hl{$\langle u \rangle:$ PIV} \\
   &   &   &   &   & \hl{$\langle \phi \rangle:$ LC \& PT} \\[0.5ex]
   
 Pohl et al.\cite{Pohl_etal2020} & \hl{I} & 14.1 & Quartz& 133 & $\langle u \rangle:$  Angled UVP  \\
   &  &  &  &  &  $\langle \phi \rangle:$  Siphon array (4) \\[0.5ex]
 
 Simmons et al.\cite{Simmons2020} & \hl{I} & 0.7 & Silt & 9.9--11$^{*5}$ & $\langle u \rangle:$  ADCPs \\
  &  &  &  &  & $\langle \phi \rangle:$  ADCPs \\
 \hline
 \multicolumn{6}{l}{MPCM: Micro-Propeller Current-Meter; UVP: Ultrasonic Velocity Profiler; }\\
 \multicolumn{6}{l}{ADV: Acoustic Doppler Profiler; ADCPs: Acoustic Doppler Current Profilers}\\
 \multicolumn{6}{l}{PIV: Particle Image Velocimetry; LC: s-beam Load Cell: PT: Pressure Transducer}\\
\caption{Summary of compiled \hl{sources}. The figures in the parentheses represent the number of \hl{measurement devices} in the array.  
$^{*1}$ Numerical simulation of Reynolds-averaged Navier-Stokes model.
$^{*2}$ Limited information \hl{about} concentration profiles \hl{reported} in this study. 
$^{*3}$ There are also experiments with sediment, but the availability of flow profiles is limited.
$^{*4}$ Original reported values of particle size from hydrometer analysis. 
$^{*5}$ $d_{50}$ from Event 1, 4, and 5 from the original source.  
$^{*6}$ Velocity meters are special equipment that are designed for their particular study.
}
\label{table:source}
\end{longtable}

The compiled published sources are listed in Table \ref{table:source}. 
In total, \hl{22 laboratory experiments}, 1 numerical simulation, and 1 direct-observation source were gathered. 
Since each source uses different tools for vertical-profile measurement, the data are compiled carefully.
For example, the velocity measurement of flume experiments can be categorized into 5 different types: micro-propeller current-meters, velocity meters, UVPs (Ultrasonic Velocity Profiler), ADVs (Acoustic Doppler Velocity probes). 
For both UVPs and ADVs, each source used a different number of probes, sampling rate, and angle of mounting. 
The vertical flow profiles reported in each source are based on interpolation and extrapolation, the methods differing between the sources. 
To perform consistent analysis, here, only the original measurement points are extracted from each sources to minimize the errors.
Then, a consistent interpolation and extrapolation method are applied to recover the full vertical profiles.
\hl{The experiments in Varjavand et al {\cite{Varjavand2015}} and Farizan et al. {\cite{Farizan2019}} are designed to analyze the effects of an obstacle to the flows.}
\hl{The experiments with obstacles are excluded from the analysis and only the experiments without obstacles are used.} 

\paragraph{Particle-size distribution analyses}
The potential error of particle size due to the different particle size analyses are carefully considered.
Settling velocity, $w_s$, plays an important role in the required work done to keep sediment in suspension, $B = gR\Phi hw_s$. 
The particle size analysis methods in the compiled \hl{sources} are either sieving combined with hydrometer method (SHM) or the laser diffraction method (LDM). SHM measures the settling velocity of the particles in a liquid. 
In SHM, the particles are assumed to be spherical and their sizes are calculated based on Stokes' law\cite{Fisher2017LDS}. 
On the other hand, in LDM, the measured particles size becomes equivalent to that of a sphere giving the same diffraction as the particles. 
LDM can process each sample faster and provide more detailed information such as the number of particles and surface area, compared with SHM\cite{Goossens2008grainsize}. 
Further, LDM provides high repeatability and reproducibility\cite{Eshel2004}. 
\par
Extensive comparison of SHM and LDM\cite{Katayama1997LDS,Vitton1997particle,Konert1997particle,lu2000accuracy,Goossens2008grainsize,DiStefano2010LDS,Al-Hashemi2021LDS,Lopez_2021LDS} has shown a large discrepancy between the measured proportion of clay-size particles. 
Although the discrepancy becomes negligible for sand-size particles in most cases, when comparing the size of the largest particles in sample of silt to clay-size particles, the error can be up to two orders of magnitude\cite{lu2000accuracy}. 
The particle-size error of two orders of magnitude can result in up to 4 order of magnitude errors of settling velocity (Eq.\eqref{eq:settling-velocity_stokes}). 
Those measurement discrepancies are attributed to the density difference and shape variation of particles\cite{Bah2009LDS}. \par
\begin{figure}[htbp]
\centering
\includegraphics{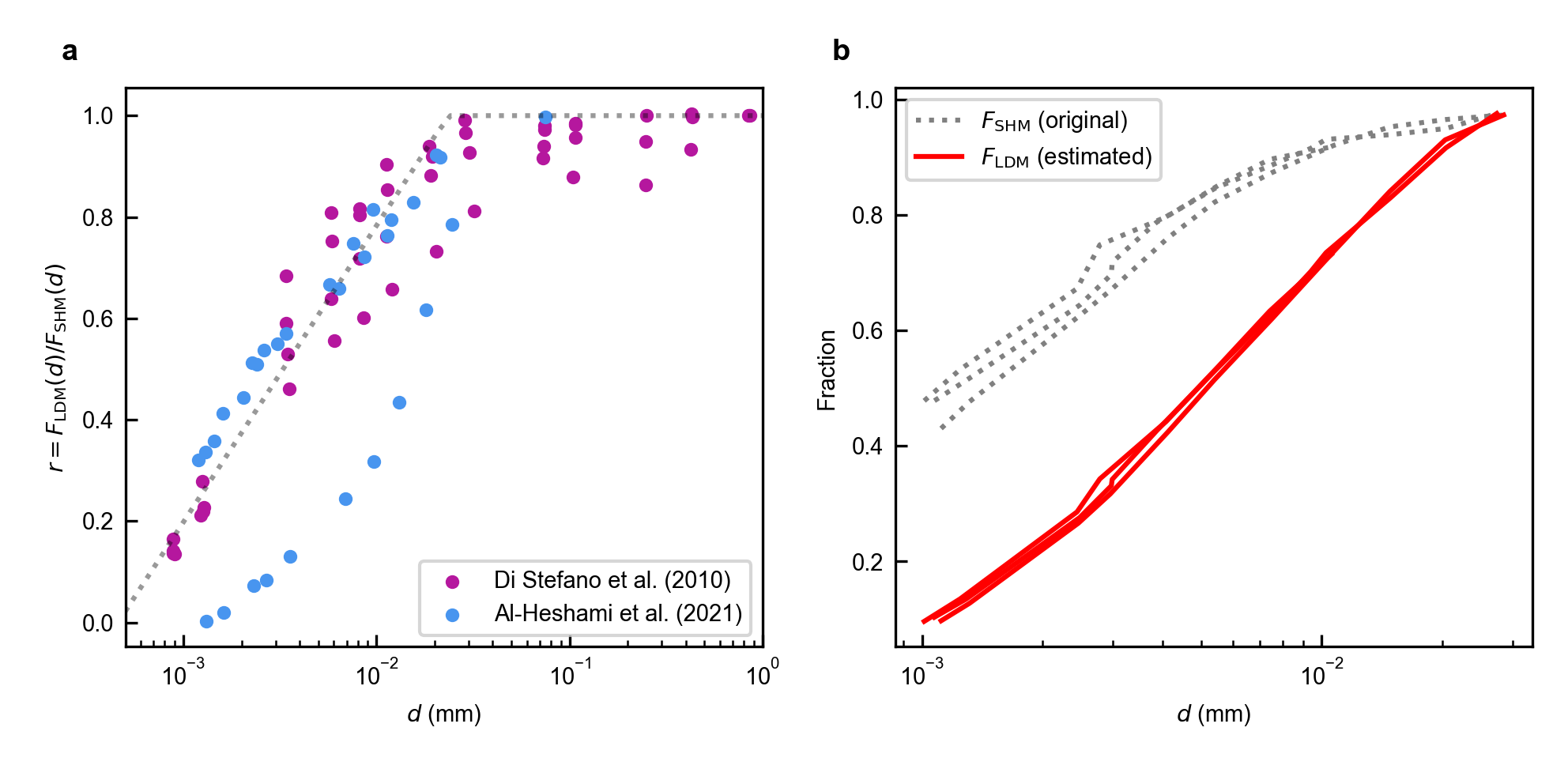}
\caption{ a) The ratio of particle sizes as measured by the Laser Diffraction method to that of the hydrometer analysis method. Gray dotted line represents the fitted curve. b) Modified particle-size distribution of Tesaker\cite{Tesaker1969uniform}. }
\label{LDM-SHM}
\end{figure}
In this study, particle size distribution from LDS is regarded as more reliable data than the SHM. 
Tesaker's experiment\cite{Tesaker1969uniform} used SHM to measure the particle size distribution of their clay material\hl{; the reported median particle size varies between 0.8--2.3 which is likely to be a significant underestimate}.
To mitigate this systematic error due to the difference of methodology, in this study, the following modification of reported particle size is conducted.
\par
Firstly, cumulative particle distribution curves of similar materials \hl{to that used by} Tesaker (Kaolinite clay) are extracted from the particle-size comparison studies between SHM and LDM\cite{DiStefano2010LDS,Al-Hashemi2021LDS}. 
Since the particle-size discrepancy between SHM and LDM could show different trends based on the type of clay minerals, only the datasets in which material is mainly composed of Kaolinite were chosen. 
The cumulative particle distribution curve, \(F(d_i)\) is defined as in
\begin{equation}
    F(d_i) = \sum_{k=0}^i f(d_k)
\end{equation}
where \(d_i\) is the characteristic particle size of the \(i^{\mathrm{th}}\) bin, \(N\) is the totall number of bins, and \(f(d_i)\) denotes the fraction of the \(i^{\mathrm{th}}\) particle size class which is given as
\begin{equation}
    f(d_i) = \frac{q(d_i)}{\sum_k q(d_k) }
\end{equation}
where \(q(d_i)\) denotes the measured weight or volume of the particles in the \(i^{\mathrm{th}}\) size bin.
Let \(r(d)\) denote the ratio of the cumulative fraction of particles measured by LDM to the one measured by SHM, such that
\begin{equation}
    F_\mathrm{LDM}(d) = r(d) F_\mathrm{SHM}(d).
    \label{eq:FSHMtoFLDM}
\end{equation}
\hl{Supplementary} Figure \ref{LDM-SHM}a shows the calculated best fit function \(r(d)\) from the compiled sources\cite{DiStefano2010LDS,Al-Hashemi2021LDS};
\(r(d)\) increases monotonically as \(d\) increases until \(r\) reaches to unity at \(d \simeq 3\times 10^{-2}\). 
\par
Here, the relationship between \(r\) and \(\log d\) for fine particles (\(d < 3\times 10^{-2}\)) is approximated by the linear curve fitting with the least-square method (Gray dotted line in \hl{Supplementary} Fig. \ref{LDM-SHM}a).
Then, assuming \(r \leq 1\), the empirical formula of \(r\) is given as
\begin{equation}
    r(d) = \min \left[ a \log_{10} d +b , 1\right]
    \label{eq:empiricalR}
\end{equation}
where the correlation coefficients are \(a = 0.58 \pm 0.04 \) and \(b = 1.95 \pm 0.09\).
Then, \(F_\mathrm{LDM}\) of Tesaker's experiments\cite{Tesaker1969uniform} is estimated from the measured \(F_\mathrm{SHM} \) by Equation \eqref{eq:FSHMtoFLDM} and \eqref{eq:empiricalR} (\hl{Supplementary} Fig. \ref{LDM-SHM}b).
Consequently, $d_{50}$ shifted from around $ 1 \times 10^{-3} $ mm to $ 4.98 \times 10^{-3} $ (\hl{Supplementary} Fig. \ref{LDM-SHM}b). 
In the analysis of \hl{the} main text, $ 4.98 \times 10^{-3} $ is used as the median particle size of Tesaker's experiments.

\paragraph{Extraction of original measurement points}
Depending on the availability of the dataset, different procedures were applied to extract data from the source.
There were two cases: i) original raw measurement data (such as UVP measurements data) is available, ii) only figures of measured data points and interpolated and extrapolated profiles are available.
In case (i), the raw data are directly used to reconstruct the flow profiles. 
In case (ii), the original measurements of vertical velocity and concentration are extracted by using a graph-read software (https://graphclick.en.softonic.com/mac).
However, some of the sources have less than three points of velocity or concentration, which is not enough to apply the extrapolation method. 
Here, assuming that the potential errors due to the choice of interpolation and extrapolation near the maximum velocity height are relatively small, a few data points are manually added near the original data points along the interpolated curve in the original source.

\paragraph{Interpolation and extrapolation of flow profile}
\begin{figure}[htbp]
\centering
\includegraphics{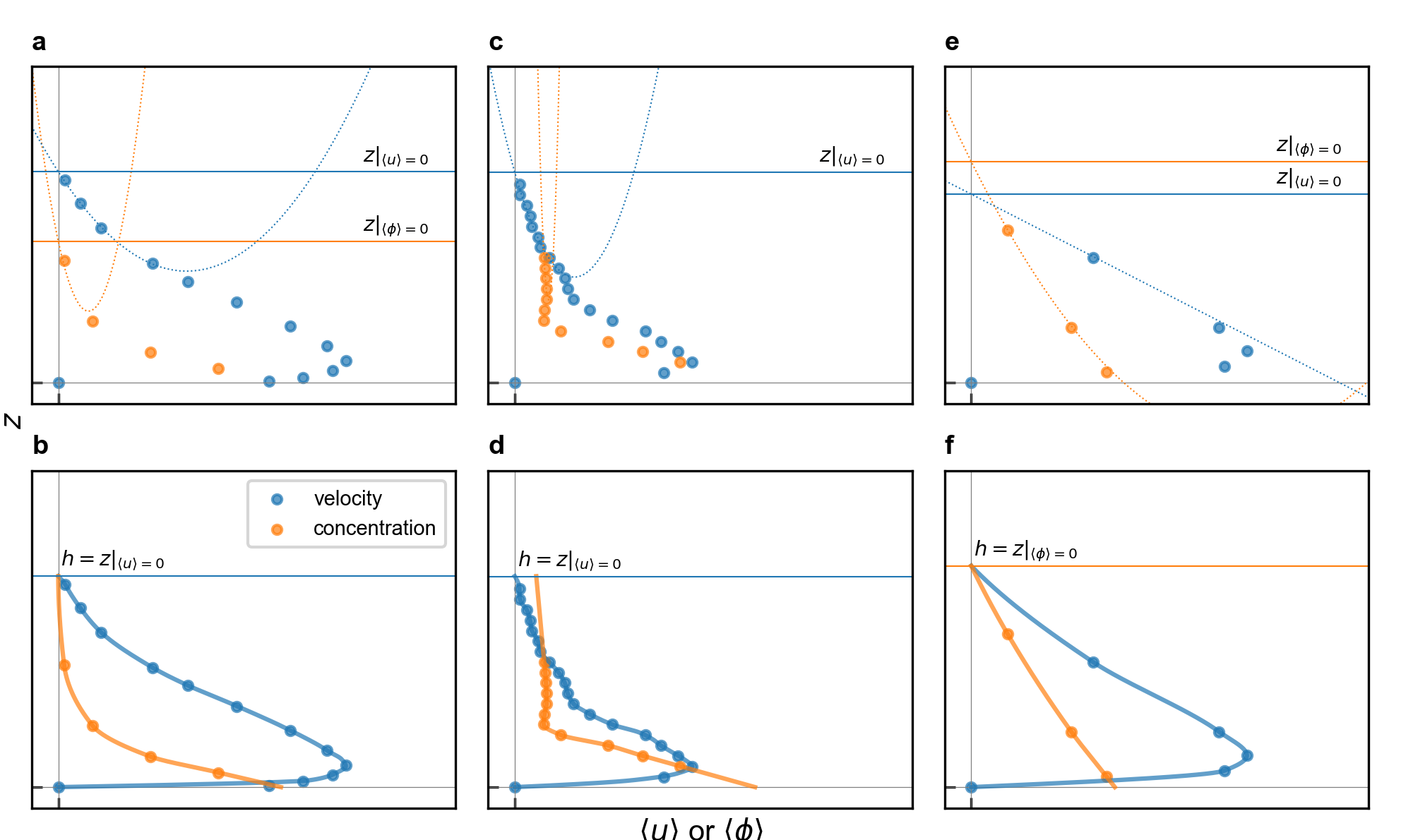}
\caption{ Schematic diagram of interpolation and extrapolation procedures for three different cases. Blue and orange markers represent the original measurement values. a) b) $z|_{\langle \phi \rangle = 0} < z_{ u,n}$. c) d) $z|_{\langle u \rangle = 0} < z|_{\langle \phi \rangle = 0}$. e) f) $z_{u,n} < z|_{\langle \phi \rangle = 0} < z|_{\langle u \rangle = 0}$. a) c) e) Schematics of quadratic curve fitting for flow height estimation. Fitted curves are represented by dotted curves. b) d) f) Schematics of resultant flow profiles. }
\label{Interpolation}
\end{figure}

\begin{figure}[htbp]
\centering
\includegraphics{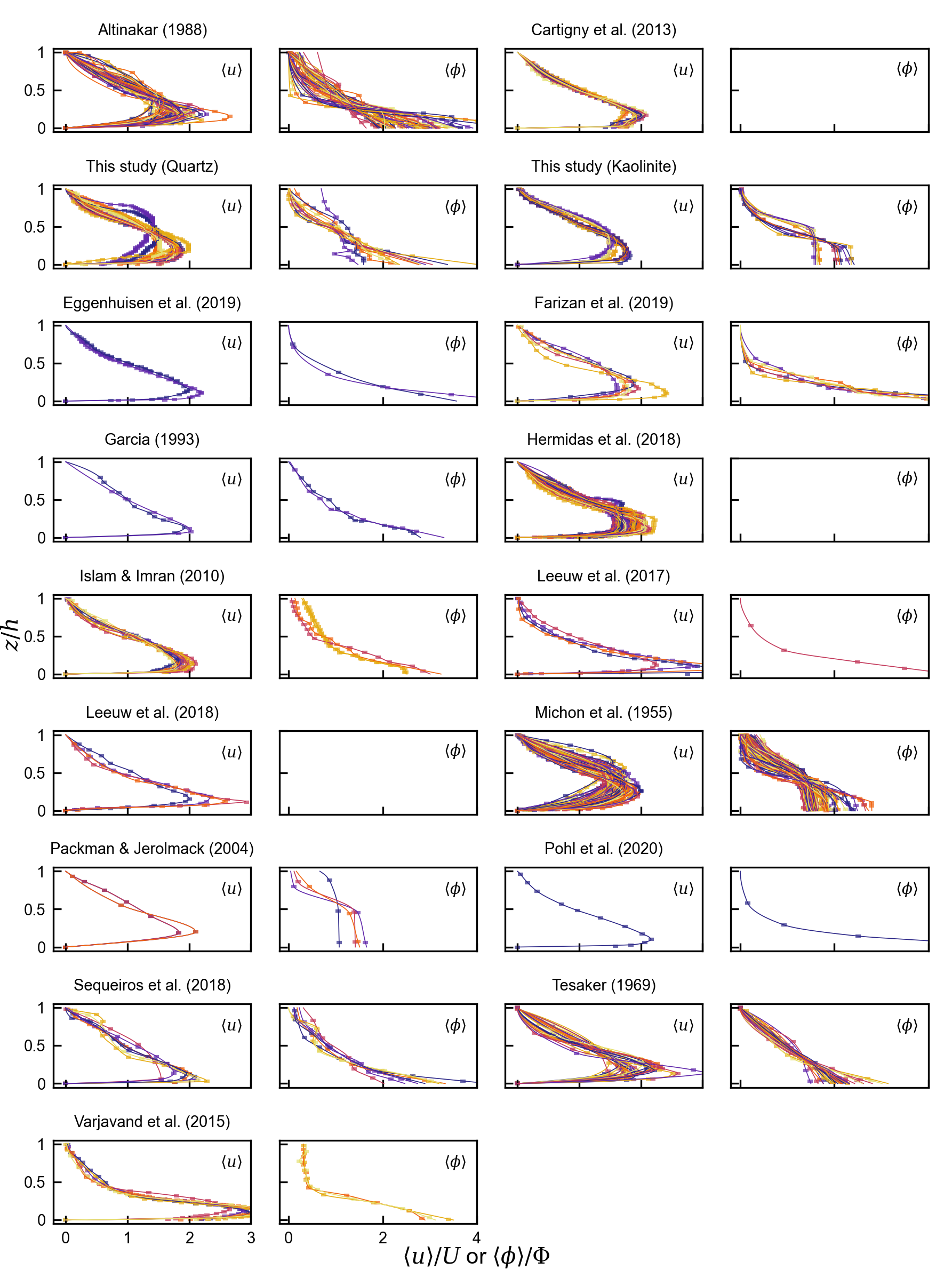}
\caption{ Interpolated and extrapolated flow profiles of the non-conservative flows from each source. For each source, velocity profiles are plotted on the left, and concentration profiles are on the right. X-axes are either flow velocity or flow concentration  which is normalized by depth-averaged values, and Y-axes are the distance from the bed normalized by flow height.}
\label{verticalprofiles01}
\end{figure}

\begin{figure}[htbp]
\centering
\includegraphics{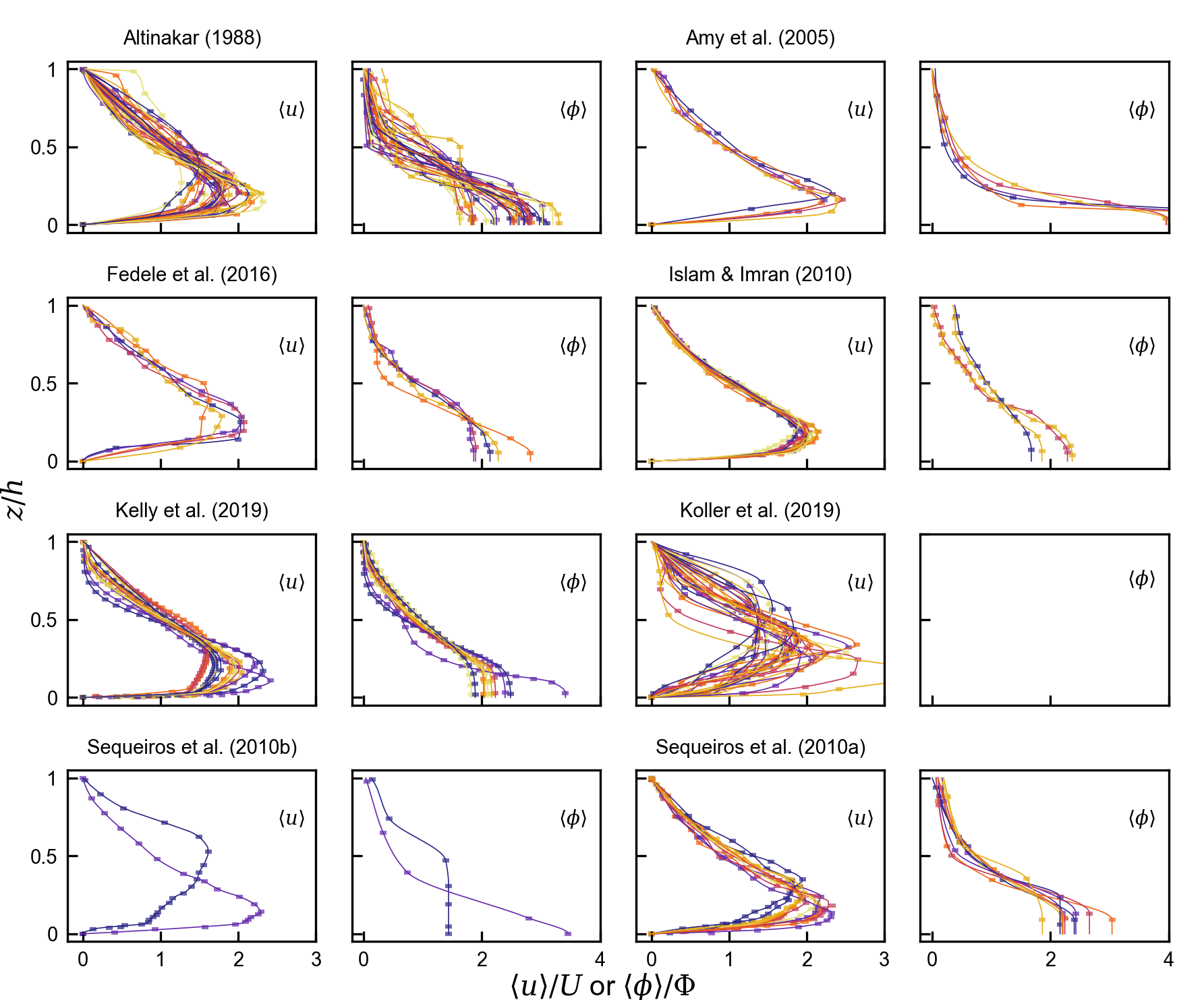}
\caption{ Interpolated and extrapolated flow profiles of the conservative flows from each source. For each source, velocity profiles are plotted on the left, and concentration profiles are on the right figure. X-axes are either flow velocity or flow concentration  which is normalized by depth-averaged values, and Y-axes are the distance from the bed normalized by flow height. }
\label{verticalprofiles02}
\end{figure}
The overall procedure can be summarized into three steps: i) data formatting, ii) estimation of flow height and near-bed concentration, and iii) interpolation and extrapolation.
Details of each step are described in this section.
\par
The first step is data formatting. 
Let $ P_{\langle u \rangle} = \left[ \mathbf{x}_{\langle u \rangle,1},\mathbf{x}_{\langle u\rangle,2},...\mathbf{x}_{\langle u\rangle,n}\right] $ and $P_{\langle\phi\rangle} = \left[ \mathbf{x}_{\langle\phi\rangle,1},\mathbf{x}_{\langle\phi\rangle,2},...\mathbf{x}_{\langle\phi\rangle,m}\right]$ denote the extracted values of velocity and concentration. 
$n$ and $m$ denote the number of extracted measurement points of velocity and concentration respectively. 
$\mathbf{x}_{\langle u\rangle,i} = (\langle u\rangle|_{z=z_{u,i}},z_{u,i} )$ and $\mathbf{x}_{\langle \phi \rangle ,i} = (\langle \phi \rangle|_{z=z_{\phi,i}},z_{\phi,i} )$ denote the $i^{\rm th}$ coordinate of extracted points. 
The data points are ordered from lowest to highest so that $z_i < z_{i+1}$. 
Thus, $ z_{u,1} $ and $ z_{\phi,1} $ are the lowest original measurement heights and  $ z_{u,n} $ and $ z_{\phi,m} $ are the highest measurement heights. 
As boundary conditions, the following relationship is assumed for velocity profiles in this study.
\begin{align}
     \langle u \rangle|_{z=0} = \langle u \rangle|_{z=h} = 0,
     \label{eq:velocity-boundary-condition}
\end{align}
To impose the lower boundary condition, an artificial data point $\mathbf{x}_{\langle u\rangle, 0} = (0,0)$ is inserted into $P_{\langle u\rangle}$.  
\par
The next step is to estimate the flow height. The estimation of the flow height is performed in two steps: i) calculation of $z|_{\langle u \rangle = 0}$ and $z|_{\langle \phi \rangle = 0}$, and then ii) evaluation of flow height based on the comparison between $z|_{\langle u \rangle = 0}$ and $z|_{\langle \phi \rangle = 0}$. 
Firstly, interpolation of $ P_{\langle u \rangle} $ and $P_{\langle \phi \rangle}$ is conducted using the PchipInterpolator function from Scipy, a Python package. 
This function provides a piecewise cubic Hermite interpolation\cite{Fritsch1984PCHIP}. 
One of the advantages of this interpolation method is that it does not overshoot even when the data is not smooth. 
Further, in this interpolation method, the original data points are always on the interpolated curve, which means that potential oversimplification is minimal.
Let $f_{\langle u \rangle}(z) $ and $f_{\langle \phi \rangle}(z)$ denote the interpolated functions. 
Then, to obtain enough data points for processing, 500 vertically uniform data points are extracted from the interpolated function,
\begin{align}
    P_{f,\langle u \rangle} = \left[ (f_{\langle u \rangle}(z_{fu,0}),z_{fu,0})... (f_{\langle u \rangle}(z_{fu,499}),z_{fu,499}) \right],\ \ 
    P_{f,\langle \phi \rangle} = \left[ (f_{\langle \phi \rangle}(z_{f\phi,0}),z_{f\phi,0}) ... (f_{\langle \phi \rangle}(z_{f\phi,499}),z_{f\phi,499}) \right].
    \label{eq:Pchip-interpolation}
\end{align}
It should be noted that the points $(f_{\langle u \rangle}(z_{fu,499}),z_{fu,499})$ and $(f_{\langle \phi \rangle}(z_{f\phi,499}),z_{f\phi,499})$ correspond to the highest original data points, $\mathbf{x}_{\langle u \rangle,n}$ and $\mathbf{x}_{\langle \phi \rangle,m}$ respectively.
Quadratic-function curve fitting is conducted against a upper part of the velocity profile to obtain $z|_{\langle u \rangle = 0}$ and $z|_{\langle \phi \rangle = 0}$ (\hl{Supplementary} Fig. \ref{Interpolation}a,c,e). 
For the velocity profile, the range of curve fitting is between $z_{fu,u_{\rm max}}$ and $z_{fu,499}$, where $z_{fu,u_{\rm max}}$ is the nearest data point to the velocity maximum. 
For the concentration profile, the range is between $z_{f\phi,249}$ and $z_{f\phi,499}$, where $z_{f\phi,249}$ denotes the data point at the middle height of interpolated region. 
The quadratic functionas, $g_u$ and $g_\phi$ are given as,
\begin{align}
    g_u(u) = a_1 \langle u \rangle^2 + a_2 \langle u \rangle + a_3, \ \ g_\phi(\phi) = b_1 \langle \phi \rangle^2 + b_2 \langle \phi \rangle + b_3.
    \label{quadratic-curve-fitting}
\end{align}
where $(a_1, a_2, a_3)$ and $(b_1,b_2,b_3)$ are the coefficients of best-fit curves estimated by the least-square method.
The conditions, $a_1 > 0$ and $b_1 > 0$ are imposed to avoid unrealistic profile shapes. 
Then, $z|_{\langle u \rangle = 0}$ and $z|_{\langle \phi \rangle = 0}$ are obtained as,
\begin{align}
    z|_{\langle u \rangle = 0} = g_u(0),\ \ z|_{\langle \phi \rangle = 0} = g_\phi(\phi_{\rm ambient}).
    \label{zero-velocity-flow-height}
\end{align}
where $\phi_{\rm ambient}$ is the concentration of the ambient fluid. Then, the flow height is given by,
\begin{align}
    h = \begin{cases}
    z|_{\langle u \rangle = 0} & \text{when $z|_{\langle u \rangle = 0} \leq z|_{\langle \phi \rangle = 0}$ \ or\ \ $ z|_{\langle \phi \rangle = 0} < z_{u,n} $ }, \\
    z|_{\langle \phi \rangle = 0} & \text{when $ z_{u,n} < z|_{\langle \phi \rangle = 0} \leq z|_{\langle u \rangle = 0}$}.
\end{cases}
    \label{flow-height-evaluation}
\end{align}
$z|_{\langle \phi \rangle = 0} < z_{u,n} $ sometimes happens when the number of data points in the density profile are inadequate. 
In this study, the original velocity measurements are regarded as more reliable than the extrapolated concentration curve. 
Thus, when $z|_{\langle \phi \rangle = 0} < z_{u,n} $ (\hl{Supplementary} Fig. \ref{Interpolation}a,b), the extrapolation of the concentration profile is discarded and $h= z|_{\langle u \rangle = 0}$ is the flow depth. 
Then, a new velocity point at the top of the flow $\mathbf{x}_{\langle u\rangle,h} = (0,h)$ is inserted into $P_u$. 
For the concentration profile, the value of the concentration at the flow height, $\mathbf{x}_{\langle \phi\rangle,h}$ is not necessarily always $\phi_{\rm ambient}$. 
When $z|_{\langle u \rangle = 0} \leq z|_{\langle \phi \rangle = 0}$, the flow concentration at the flow height is estiamted by a piecewise cubic Hermite interpolation of original measurement points with $(0,z|_{\langle \phi \rangle = 0})$. Let $f_\phi^\prime (z)$ denote the interpolated function. Then, the flow concentration at the flow height is given as $f_{\langle \phi \rangle}^\prime(z=h)$. \par
The final step is the interpolation and extrapolation of the rest of the profiles. Extrapolation is required to estimate the near-bed concentration, $\langle \phi \rangle_{\rm bed}$. For the non-conservative currents, the first-order polynomial curve fit is conducted against the lowest two data points. For the conservative flow, the lowest original measurement point $\langle \phi\rangle|_{z = z_{\phi,0}}$ is used as the near-bed concentration.
Then, combining the original measurements points, near-bed points, and flow height points, the following completed data lists were obtained,
\begin{align}
    P_{\langle u \rangle}^\prime = \left[ \mathbf{x}_{\langle u \rangle,{\rm bed} }, \mathbf{x}_{\langle u \rangle,1},...\mathbf{x}_{\langle u\rangle,n},\mathbf{x}_{\langle u\rangle,h} \right],\ \ 
    P_{\langle\phi\rangle}^\prime = \left[\mathbf{x}_{\langle \phi\rangle,{\rm bed}}, \mathbf{x}_{\langle\phi\rangle,1},...\mathbf{x}_{\langle\phi\rangle,m},\mathbf{x}_{\langle \phi\rangle,h}\right].
    \label{eq:full-data-point-list}
\end{align}
Finally, by conducting piecewise cubic Hermite interpolation against $P_{\langle u \rangle}^\prime$ and $P_{\langle\phi\rangle}^\prime$, the full profile $\langle u \rangle(z)$ and $\langle \phi \rangle(z)$ were obtained. The full profiles for each source obtained by the methodology described here are plotted in \hl{Supplementary} Figure \ref{verticalprofiles01}, \ref{verticalprofiles02}.

\paragraph{Turbidity currents from Congo canyon}
\begin{figure}[htbp]
\centering
\boxhl{
\includegraphics{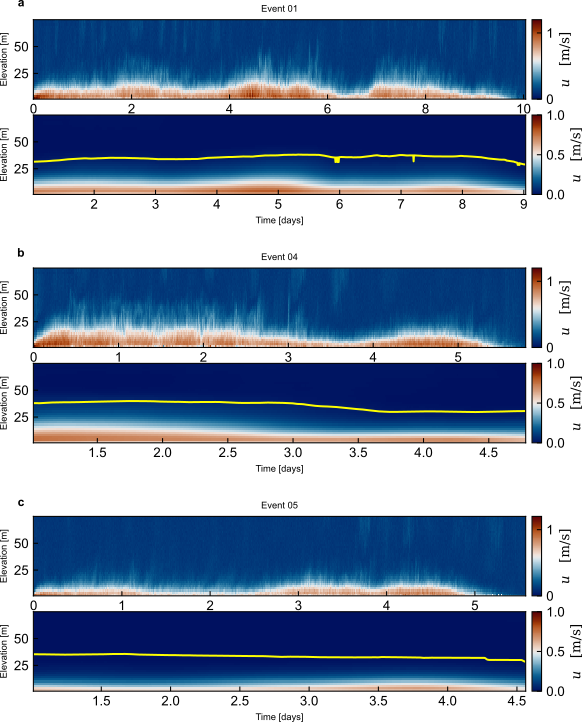}
}
\caption{\hl{Time series of original and moving-averaged velocity field of the turbidity currents in the Congo system. Yellow lines are the estimated flow height.}}
\label{TimeseriesCongo}
\end{figure}

\begin{figure}[htbp]
\centering
\boxhl{
\includegraphics{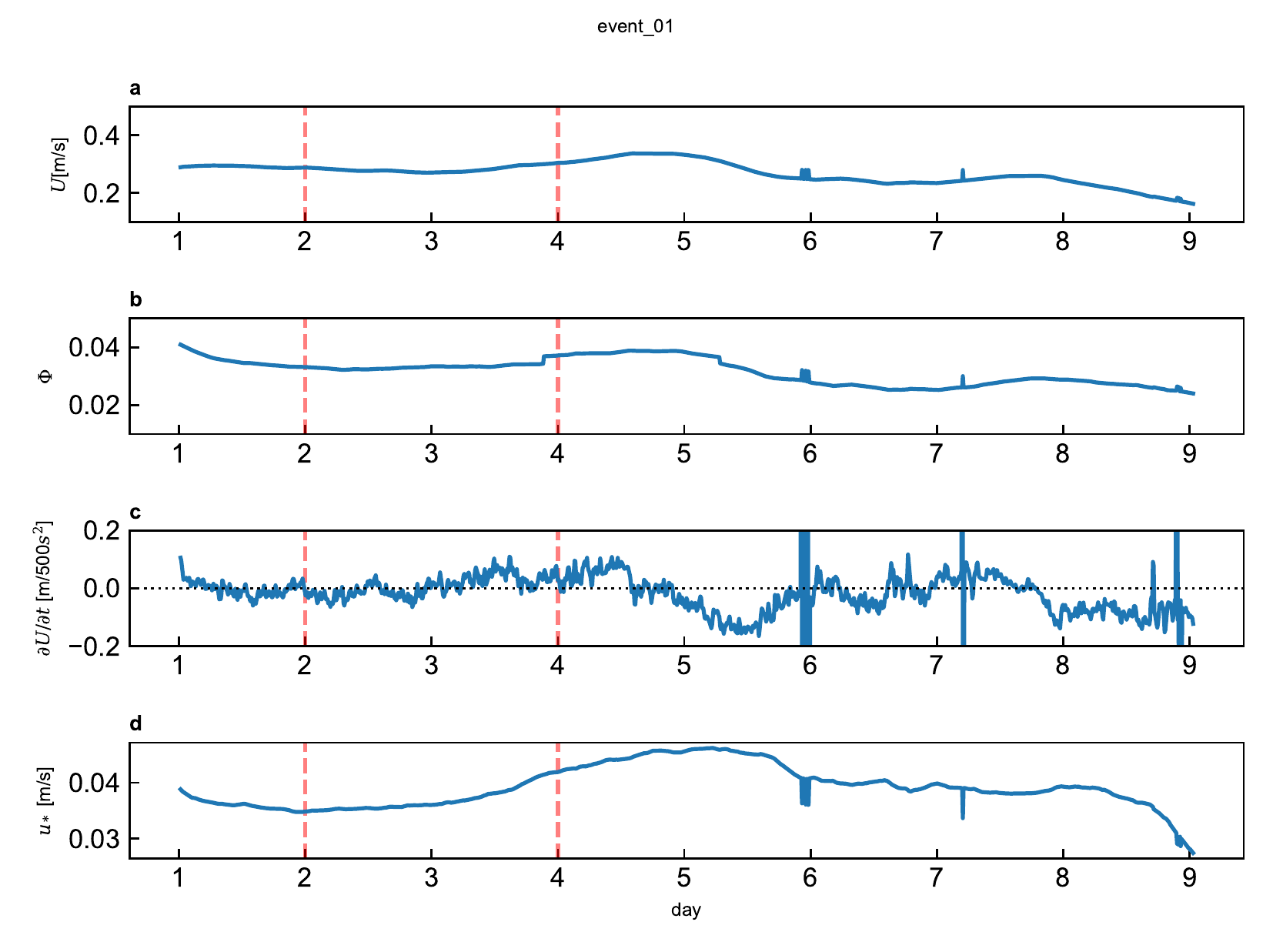}
}
\caption{\hl{Depth-averaged flow parameters of Event 01 in Congo canyon. a) Depth-averaged flow velocity, b) depth-averaged flow concentration, c) time gradient of depth-averaged flow velocity [m/500$s^2$] and, d) estimated shear velocity. The selected time interval is indicated by red dashed lines.}}
\label{event01}
\end{figure}

\begin{figure}[htbp]
\centering
\boxhl{
\includegraphics{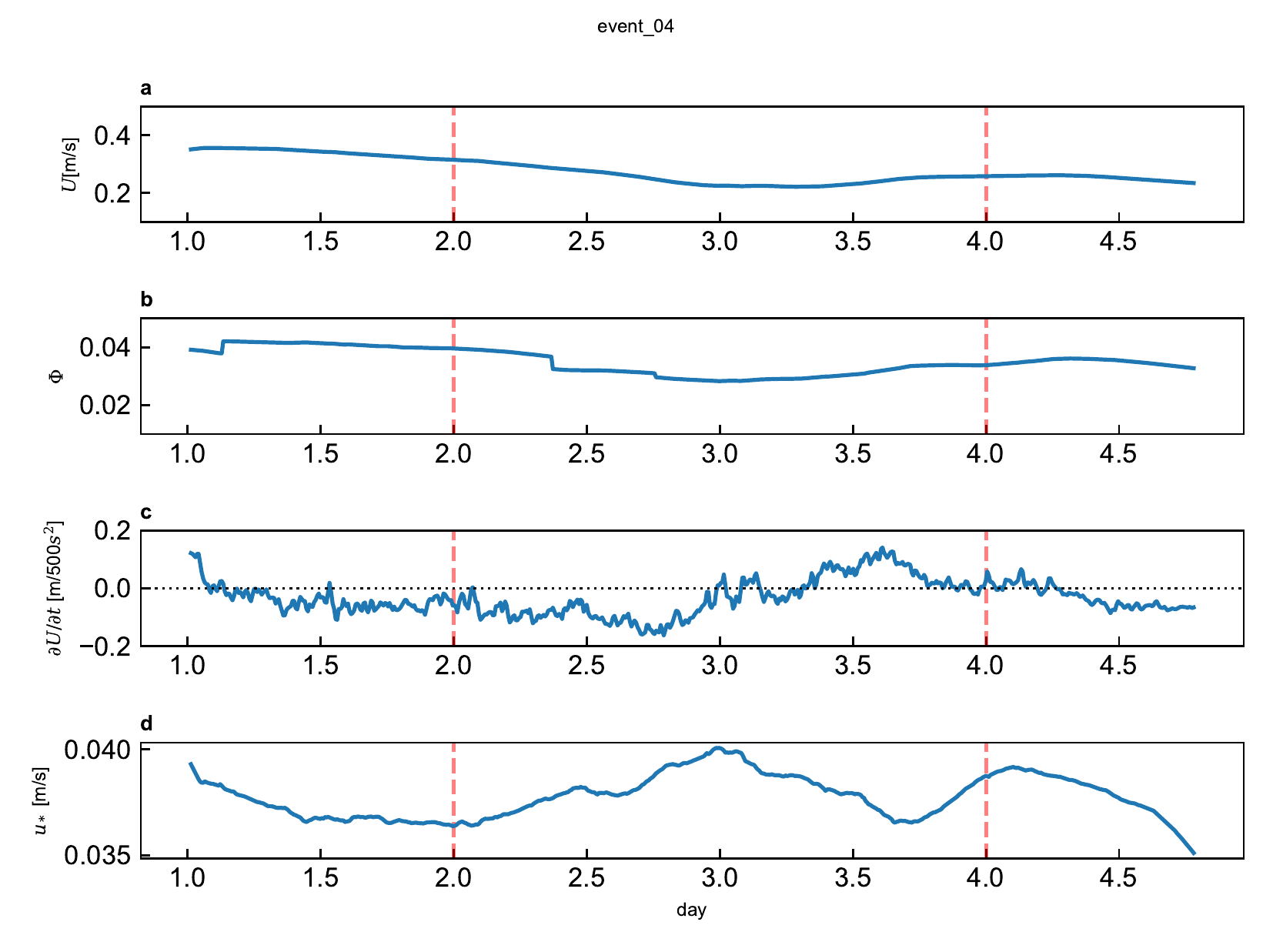}
}
\caption{\hl{Depth-averaged flow parameters of Event 04 in Congo canyon. a) Depth-averaged flow velocity, b) depth-averaged flow concentration, c) time gradient of depth-averaged flow velocity [m/500$s^2$] and, d) estimated shear velocity. The selected time interval is indicated by red dashed lines.}}
\label{event04}
\end{figure}

\begin{figure}[htbp]
\centering
\boxhl{
\includegraphics{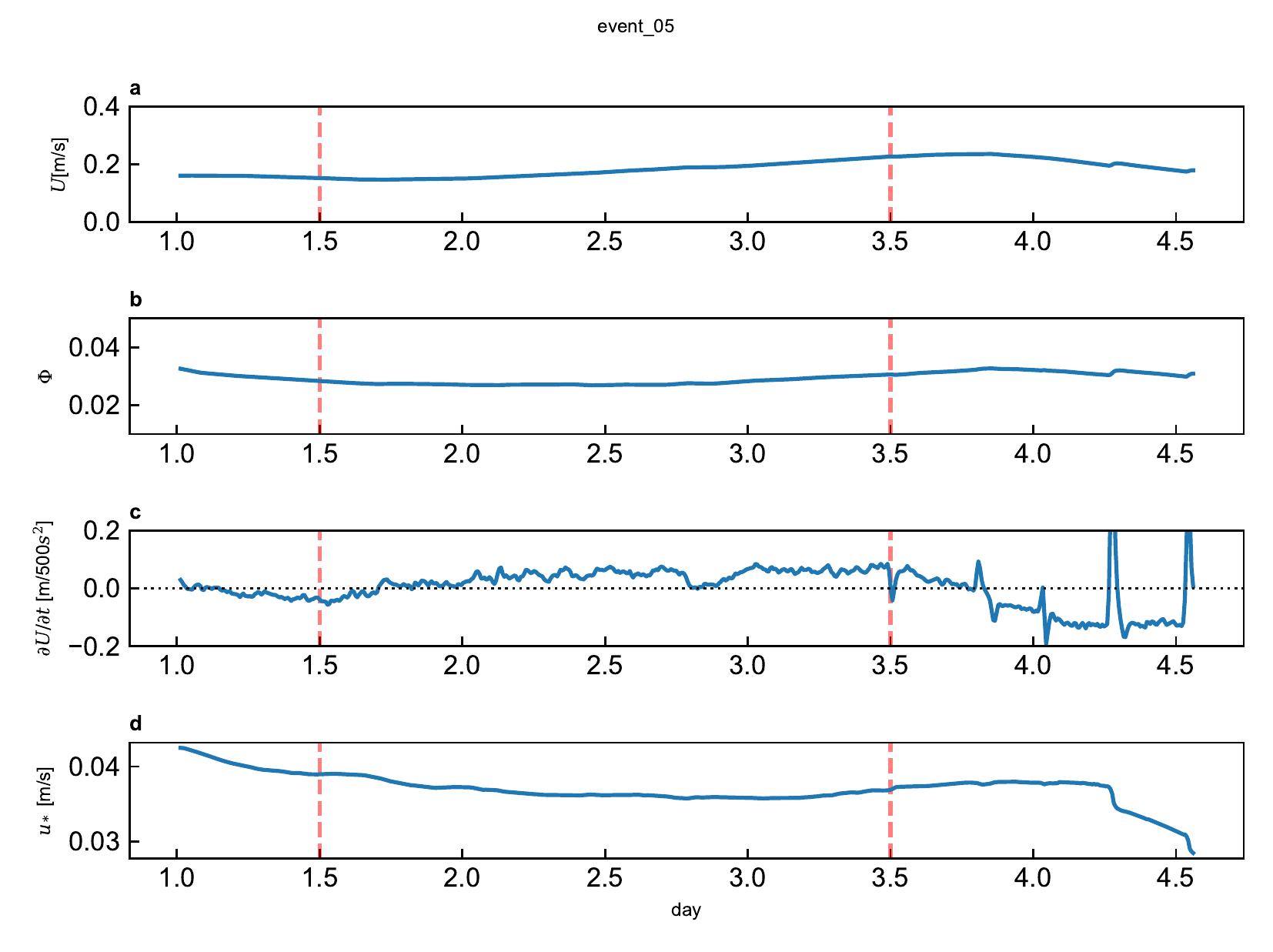}
}
\caption{\hl{Depth-averaged flow parameters of Event 05 in Congo canyon. a) Depth-averaged flow velocity, b) depth-averaged flow concentration, c) time gradient of depth-averaged flow velocity [m/500$s^2$] and, d) estimated shear velocity. The selected time interval is indicated by red dashed lines.}}
\label{event05}
\end{figure}

The real-world dataset from Congo canyon is extracted from the \hl{publication by Simmmons et al.}\cite{Simmons2020}. 
\hl{Here}, flow velocity and concentration structures are measured by Acoustic Doppler Current Profilers (ADCPs). 
In this study, to investigate the flow dynamics of quasi-equilibrium natural-scale turbidity currents, relatively long-duration (more than 5 days) events are chosen (Event 01, 04, and 05). 
Some long-duration events in which velocity and concentration fields drastically changed during the flow event (such as Events 8, 9, and 10 from the original source) are excluded.
\par
\hl{Firstly, the moving averages of the velocity and concentration time series are calculated from each event (Supplementary Fig. {\ref{TimeseriesCongo}}).}
\hl{Then, the depth-averaged flow parameters, (flow velocity and concentration), are calculated from each averaged profile (Supplementary Fig. {\ref{event01}}--{\ref{event05}}).}
\hl{Based on the gradient in time of the depth-averaged flow velocity, intervals of time for data compilation are selected so that the strongly accelerated or decelerated regions are excluded.}
\hl{Finally, from the selected window, mean depth-averaged flow parameters are calculated.}
To remove the background noise from each flow, the time-average background velocity with in the selected window is calculated for each run.
Then \hl{velocity values lower than the background noise are excluded in the interpolation and extrapolation of profiles.}
\hl{The velocity relative to} the time-averaged background velocity is extracted as the flow velocity measurements. 
\par
The original concentration profiles include some artifacts\cite{Simmons2020}. 
In events 1 and 4, sometimes concentration values suddenly increase near the interface with the ambient. 
Simmons et al.\cite{Simmons2020} assumed that this type of artefact is related to the backscatter from turbulent microstructure associated with gradients in either density, temperature, or salinity\cite{Lavery2003}. 
To exclude the artefacts, the peak value of the concentration near the upper flow interface is measured from Event 1 and 4.
Then, in the upper half of the flow, the concentration lower than this peak is omitted; only the portion of the concentration data which is of higher concentration than the artefact is used for interpolation and extrapolation.

\par
\paragraph{Estimation of depth-averaged concentration}
\begin{figure}[htbp]
\centering
\includegraphics{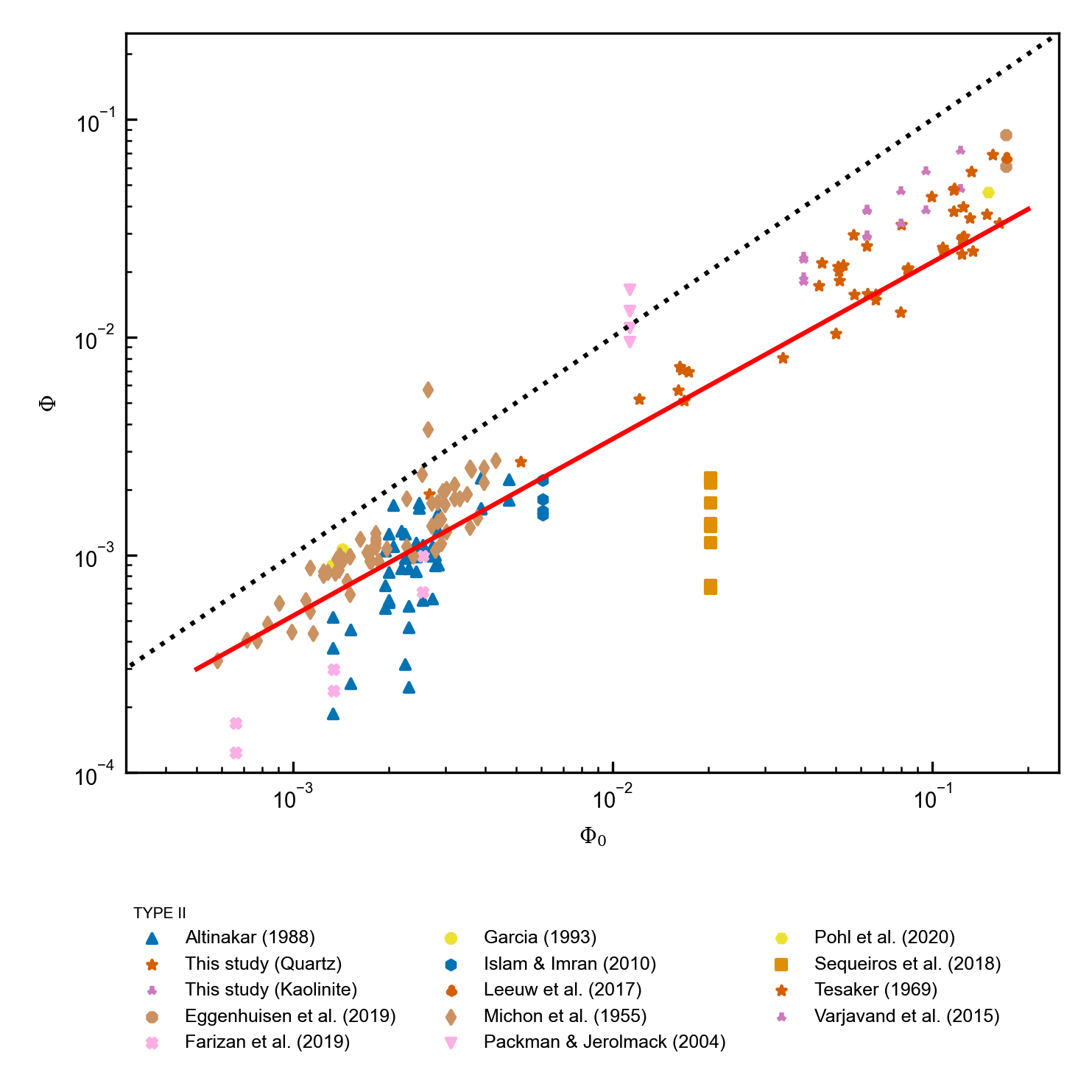}
\caption{Log-log plot of layer-averaged flow concentration and initial concentration of the sediment-water mixture in the mixing tank of each experiment. The regression line (gray dotted line) was fitted to logarithmic values of $\Phi_0$ and $\Phi$. The coefficient of determination is $R = 0.88$. }
\label{CzeroRatio}
\end{figure}
To estimate the depth-averaged sediment concentration, $\Phi$, for the experiments in which the vertical concentration profiles are not available (TYPE III, see Table \ref{table:source}), the following empirical relationship between $\Phi$ and the initial sediment concentration in the mixing tank, $\Phi_0$\hl{,} are deduced, using orthogonal distance regression, from TYPE II data where both parameters are available (\hl{Supplementary} Fig. \ref{CzeroRatio}).
Then, $\Phi$ of TYPE III data is assumed to follow the obtained empirical formula,
\begin{align}
    \Phi \simeq 0.30 \times {\Phi_0}^{0.94}
\end{align}

\paragraph{Drag coefficient}
\begin{figure}[htbp]
\centering
\boxhl{\includegraphics{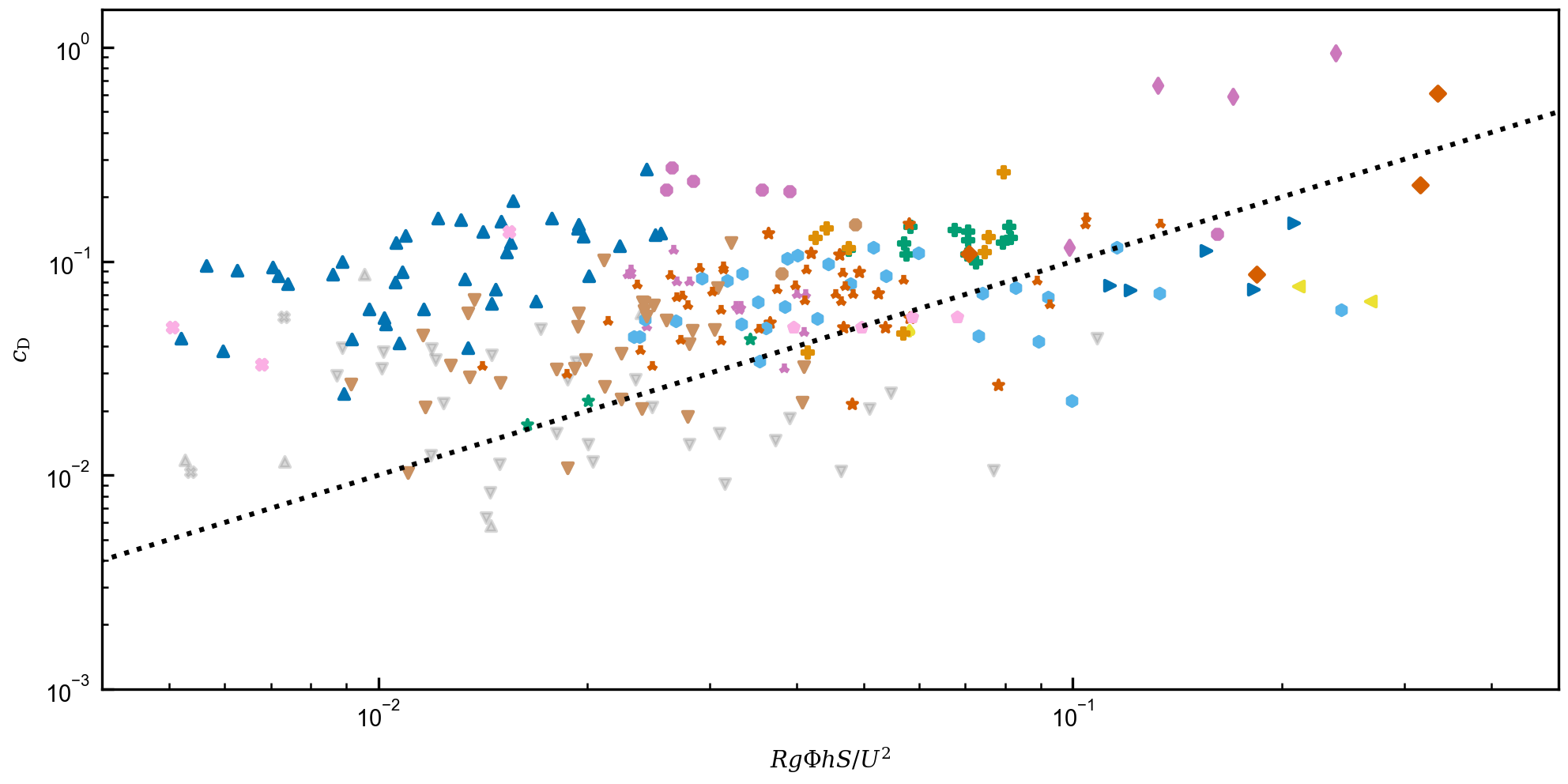}}
\caption{\hl{Log-log plot of the estimated drag coefficient (Eq. {\ref{eq:ustarfromgradient}}--{\ref{eq:skindrag}}) verses the top-hat drag coefficient{\cite{Parker1986}} (Eq. {\ref{eq:CD_tophat}}).}
\hl{The black dotted line represents the ideal linear line ($C_{\mathrm{D}}=Rg\Phi hS/U^2$). The gray points indicate that those data points are excluded by the near-equilibrium criteria (Fig. 2 in the main text).}
}
\label{fig:DragCoefficient}
\end{figure}
\hl{In this study the drag coefficient, $C_\mathrm{D}$ is estimated from the velocity gradient (Eq. {\ref{eq:ustarfromgradient}}--{\ref{eq:skindrag}}).
On the other hand, when a flow is in equilibrium, top-hat (unstratified) shallow water models{\cite{Parker1986}} give
}
\begin{align}
    C_\mathrm{D} = \frac{Rg\Phi hS}{U^2},
    \label{eq:CD_tophat}
\end{align}
\hl{where $S$ denotes the energy slope.}
\hl{Here, we compare the drag coefficient values estimated from the velocity gradient (Eq. {\ref{eq:skindrag}}) and those values based on the top-hat equilibrium assumption (Eq. {\ref{eq:CD_tophat}}), see Supplementary Figure {\ref{fig:DragCoefficient}}.}
\hl{The top-hat drag coefficient shows relatively better correlation with velocity-profile-based drag coefficient in the relatively large drag coefficient region ($Rg\Phi hS/U^2 \gtrsim 0.05$).
While the flows with the low top-hat drag coefficient ($Rg\Phi hS/U^2 \lesssim 0.05$) show a relatively poor correlation with the velocity-profile-based drag coefficient. }
\hl{This serves to further stress the main conclusion, that the top-hat model does not accurately capture the dynamics of gravity currents.}

\section{Flume Experiments}
\begin{figure}[htbp]
\centering
\includegraphics{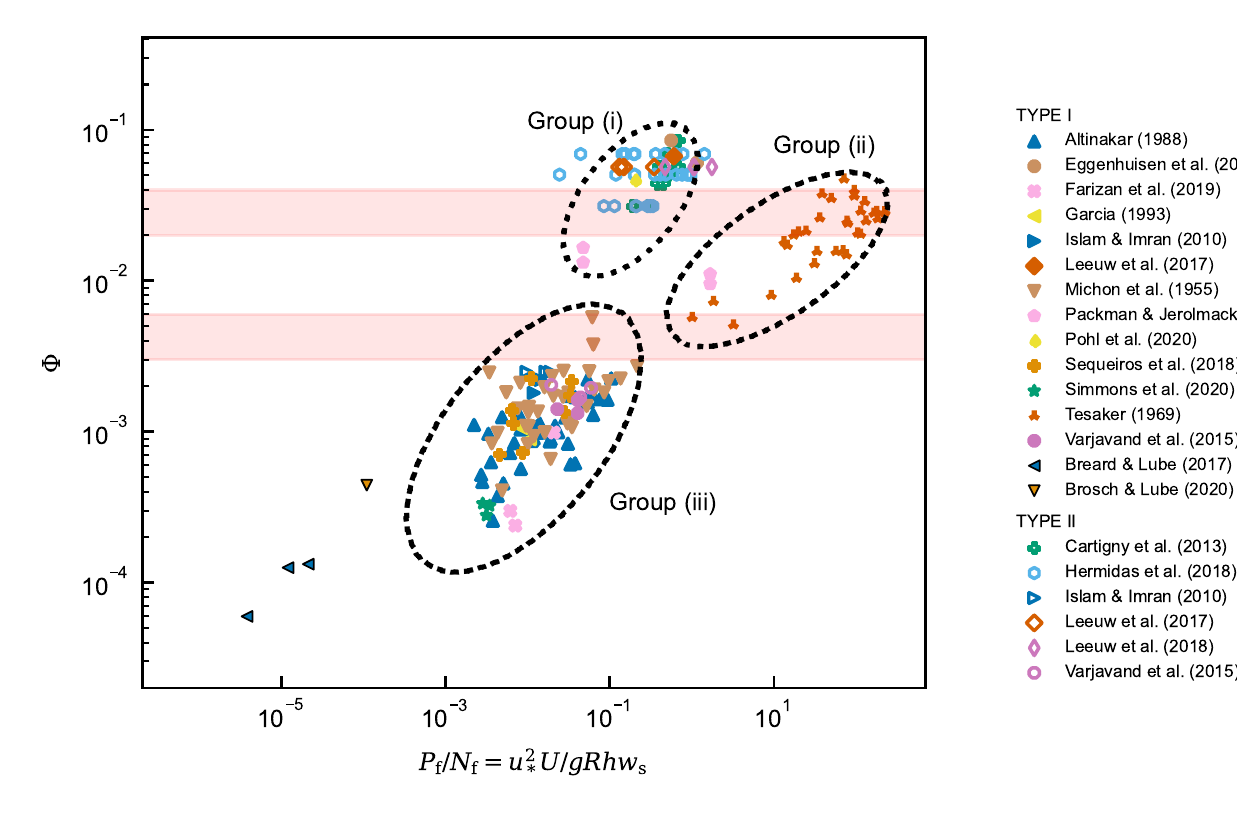}
\caption{Flow power plot with target flow concentration regions (red shaded area) of the flume experiments. Part of the experiments was aimed at filling the gap between groups of data points (i) and (ii). The rest of the experiments were aimed to supply the data between the group (ii) and (iii).}
\label{Exp-setting}
\end{figure}
The experiments are carried out in the flume laboratory in the Total Environmental Simulator at The DEEP. 
The flow properties (Table 1 in the main text) are chosen to fill the gaps between the data points from the compiled sources (\hl{Supplementary} Fig. \ref{Exp-setting}). 
A schematic and description of the experimental settings is given in the main text (Fig. 6 in the main text).
Here, the preparation of the sediment-water mixture\hl{,} and \hl{the} bed profile measurement which is conducted to measure the aggradation rate\hl{,} are elaborated.
\paragraph{Preparation of sediment-water mixture}
A large corn-shape mixing tank is built and used to generate turbidity currents (in total 1.36 m$^3$ including the region below the mixer which cannot be used; the total effective volume for the experimental run above the mixer is approximately 0.97 m$^3$).
To keep the concentration vertically uniform, two different mixing systems are established: i) an electrical mixer with a set of 350 mm diameter impellers is inserted from the top of the tank which generates a strong vortex within the tank, and 2) a re-circulation piping system with a pump is connected from the bottom of the tank to the top of the tank which generates vertical circulation.
Another piping system is connected with a slurry pump (Ebara DWO 300) from the fit on the sidewall of the tank to the flow diffuser inserted upstream of the perspex straight channel.
The (pumped) input rate is tested by adjusting the in-line valve, reading the flow rate from an electromagnetic flowmeter during test runs. 
The sediment concentration in the mixing tank and in each flow are calculated from the wet and dry weight of collected samples. Finally, the dry sample was analyzed by a laser particle sizer (Malvern Mastersizer) to estimate the detailed particle size distribution.

\paragraph{Bed profile}
Aggradation rate, for flow equilibrium, and bed depth is monitored by using the bed profiler from a single ADV (Acoustic Doppler Velocimeter) suspended above the flow. In addition to the ADV, 4 ultrasonic sensors (URSs) to monitor the aggradation rate are mounted. To reduce mobile bedload, and ADV measurement noise, experiments start with the unerodible bed of the Perspex channel. The maximum aggradation rate of the well-developed flow body observed in our experiments is 0.1 mm/s in run 11. It should be noted that for high concentrated runs (experiments 01--07 and 13--17, see Table 2 in the main text) both ADV and URSs failed to monitor the bed height during the flow events. For those runs, the aggradation rate is monitored from the videos that were captured by high-resolution GoPro cameras. As result, for those high concentration runs, deposition only occurred at the very end of the flow event (the tail of the flow) but we observed almost no deposition from the head and body of each flow where the velocity and density measurement are conducted. 

\section{Flow power plot}
\hl{From the flow power theory {\cite{velikanov1954,Bagnold1966approach}} and the autosuspension criteria {\cite{Bagnold1962}}, the following proportionality is assumed}
\begin{align}
\mathhl{
    P_\mathrm{shear} \propto B_\mathrm{turb}, \qquad \text{where} \qquad P_\mathrm{shear} = -\int_0^\infty \ab{u^\prime w^\prime} 
    \pfrac{\ab{u}}{z}\ \drm{z}\qquad \text{and} \qquad B_\mathrm{turb} = \int_0^\infty \Tilde{B}\ \drm{z} = \int_0^\infty Rg \ab{w^\prime \phi^\prime}\ \drm{z} , }
    \label{eq:flowpower}
\end{align}
\hl{where $P$ denotes the shear production of the mean flow, and the buoyancy production term, $B$, denotes the work done by turbulence to keep sediment in suspension.} 
\hl{For the fluvial flows, assuming $B_\mathrm{f}=\Phi N_\mathrm{f}=Rg\Phi hw_\mathrm{s}$ and the direct proportionality between total production, $P_\mathrm{loss}$ and the log-law production, $P_\mathrm{f}={u_*}^2U$,}
\begin{align}
\mathhl{
    \Phi \propto \frac{P_\mathrm{f}}{N_\mathrm{f}} = \frac{u_*^2U}{Rghw_\mathrm{s}}, }
    \label{eq:flowpower_OC}
\end{align}
\hl{is obtained.}
\par
\begin{figure}
    \centering
    \boxhl{\includegraphics{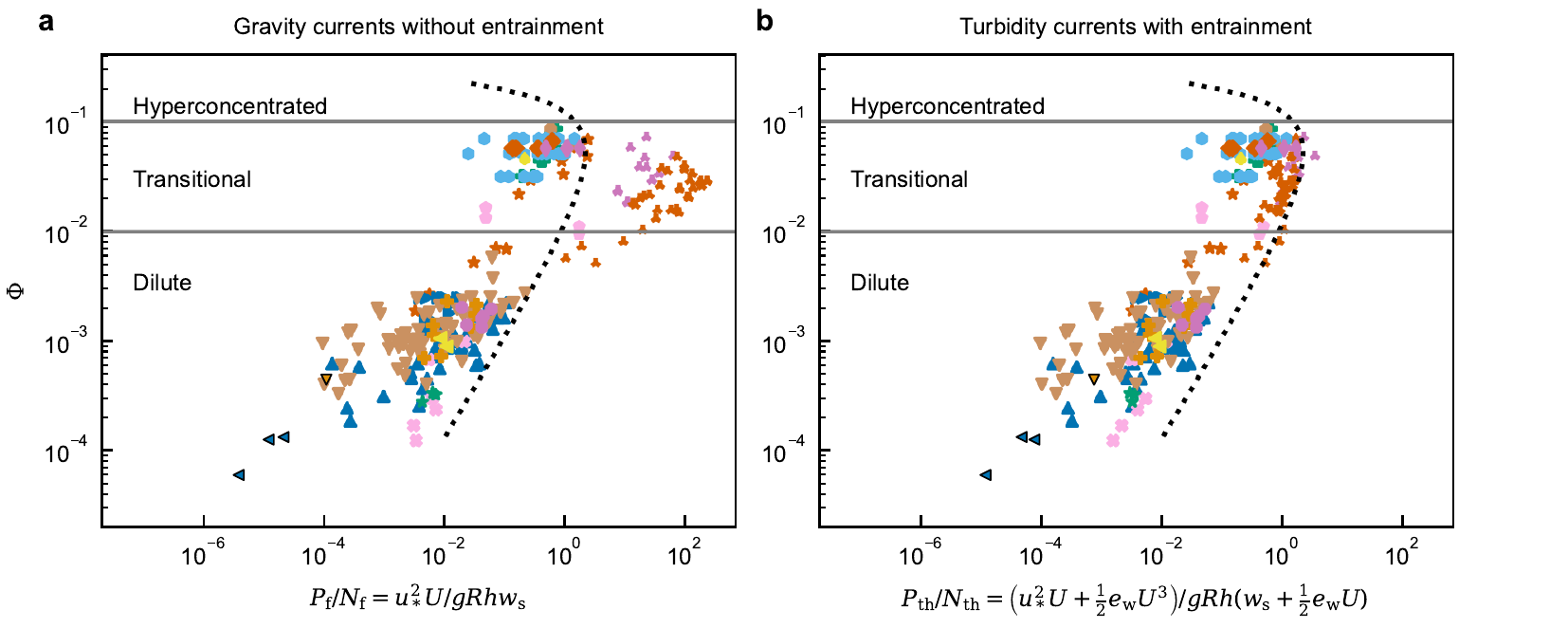}}
    \caption{\hl{Sediment transport capacity for gravity currents without data reduction.
Sediment concentration, $\Phi$, versus dimensionless flow power, $P_\mathrm{f}/N_\mathrm{f}$ for: 
a) all gravity current data with the log-law production term, $P_\mathrm{f}$;
b) all gravity current data with the top-hat production term, $P_\mathrm{th}$.
Throughout, dilute, transitional and hyperconcentrated regimes are separated by gray solid lines.
Parametric correlations of concentration and dimensionless flow power are depicted by black dotted curves.
Symbol shapes as per Figure 1 in the main text.}
}
    \label{fig:fpplotall}
\end{figure}

\paragraph{Laboratory-scale turbidity currents data}
\hl{There are few data points in the dilute regime around $10^{-1} < P_\mathrm{f}/B_\mathrm{f} < 10^{0}$}.
\hl{This is not because the data is excluded due to the introduced equilibrium criteria (Fig. 2 in the main text) but simply the long-duration experiments of turbidity currents with flow concentration, $\Phi \sim 0.5 \%$ are limited (Supplementary Fig. {\ref{fig:fpplotall}}).}
\hl{Our flume experiments partially fill this data gap by adding three runs within the target concentration range.}

\paragraph{Pyroclastic density currents}
\hl{The recent laboratory experiments{\cite{BreardLube2017,BroschLube2020}} of pyroclastic density currents (PDCs) are gathered (see Table. {\ref{table:source}}).}
\hl{The same methodology of data compilation for turbidity currents is applied for PDCs.}
\hl{For the entrainment rate of air of pyroclastic density currents, the empirical equation{\cite{Dellino2019pdc}}}
\begin{align}
    \mathhl{e_w = 0.21\mathrm{Ri_o}^{-1.1}}
\end{align}
\hl{is used, where $\mathrm{Ri_o} = g^\prime h \cos{\theta}/U^2$ is the overall Richardson number, $g^\prime$ is the reduced gravity, and $\theta$ denotes the angle of slope. }
\paragraph{Fluvial flows}
\hl{The fluvial data from natural-scale rivers{\cite{Nordin1963vertical,Wan1994hyper}}} and laboratory-scale experiments{\cite{Guy1966summary,AshidaOkabe1982OC,CellinoGraf1999OC,GrafCellino2002OC,einstein1955effects,Coleman1986OC,lyn_1988,Vanoni1946OC,VanoniNomicos1960OC,Brooks1954laboratory}} are gathered (Fig. 3 \hl{in the main text} and \hl{Supplementary} Fig. \ref{fig:fpplotall}a).
\hl{The Yellow river data {\cite{Wan1994hyper}} are gathered from the figure from {\cite{vanMaren2007}}, using a graph reading software.}
\hl{Since the shear velocity values and raw velocity profiles of Yellow River data points are not available in the original source, the average skin drag coefficient, $C_\mathrm{D}$ of Yellow River data is estimated from the recent velocity measurement data{\cite{Moodie2022YR}}. }
The reported flow parameters from each source such as flow velocity, volumetric flow concentration, and median particle size are directly used for the calculation.
\hl{For the detailed data compilation methodology for the rest of the data, see Dorrell et al.{\cite{Dorrell2018equilibrium}}}.
\begin{longtable}{ l | l  } 
 \hline
 source & site \\ [0.5ex] 
 \hline 
 Vanoni \cite{Vanoni1946OC} & Laboratory exp. \\
 Brooks \cite{Brooks1954laboratory} & Laboratory exp. \\
 Einstein \& Chien \cite{einstein1955effects} & Laboratory exp. \\
  Vanoni \& Nomicos \cite{VanoniNomicos1960OC} & Laboratory exp. \\
 Nordin \& Dempster \cite{Nordin1963vertical} & Rio Grande \\
 Guy et al. \cite{Guy1966summary} & Laboratory exp. \\
 Coleman \cite{Coleman1986OC} & Laboratory exp. \\
 Lyn \cite{lyn_1988} & Laboratory exp. \\
 Wan \& Wang \cite{Wan1994hyper} & Yellow River \\
 Ashida \& Okabe \cite{AshidaOkabe1982OC} & Laboratory exp.\\
 Cellino \& Graf \cite{CellinoGraf1999OC} & Laboratory exp.\\
 Graf \& Cellino \cite{GrafCellino2002OC} & Laboratory exp.\\
 \hline
\caption{Summary of the compiled fluvial source.}
\label{table:sourceOC}
\end{longtable}

\section{Curve Fitting}
\label{sec:curvefit}
In the main text, two types of regression analyses are conducted. For the simple linear regression, Orthogonal Distance Regression (ODR) method\cite{BoggsODR1990} is used.
\hl{While Ordinary Least Squares (OLS) treat the data for explanatory variables ($x$-axis) as without error, minimising the distance from the response variable ($y$-axis) to the fitted curve, ODR accounts equally for errors of both explanatory and response variables, minimising the orthogonal distance between each data point and the fitted curve.}
\hl{All explanatory variables in the linear regressions in this study are expected to exhibit error due to (for example) the limitation of measurement tools such as UVP or siphon arrays. Thus, ODR is more suitable than OLS.}
\hl{In Table 1 in the main text, the standard error of each estimated parameter is given, which is calculated from the square root of the corresponding diagonal term in the parameter covariance matrix.}
\hl{To infer the goodness of fit of ODR, the coefficient of determination, $R^2$ is introduced.}
\hl{$R^2$ is calculated from the sum of squares of residuals, $\mathrm{RSS}$ and the total sum of squares, $\mathrm{TSS}$ as,}
\begin{align}
    R^2 = 1 - \frac{\mathrm{RSS}}{\mathrm{TSS}},
\end{align}
\hl{where $\mathrm{RSS}$ and $\mathrm{TSS}$ for ODR is calculated as,}
\begin{align}
    \mathrm{RSS} &= \sum_i d(x_i,y_i,f(\bm{\beta}))^2\\
    \mathrm{TSS} &= \sum_i (y_i - \overline{y})^2,
\end{align}
\hl{where $(x_i,y_i)$ denotes the coordinate of the $i^\mathrm{th}$ observed data, $d(x_i,y_i,f)^2$ denotes the squared orthogonal distance between $(x_i,y_i)$ and the fitted curve $f$, and $\overline{y}$ denotes the mean value of $y_i$. }

\paragraph{Flow power plot}
To fit a curve to the data in figure 3a in the main text, we use the least squares fit for a parametric curve fitting. To simplify notation, here we will use the notation \hl{$x = \log_{10}(P_f/Rghw_s)$}, $y=\log_{10}(\Phi)$, so that the goal is to fit a curve $(x(t),y(t))$ to the given set of data points $(x_i,y_i)$. The curves are written as polynomials in the parameter $t$, and we choose to take $x$ as a quadratic in $t$ and $y$ as a cubic in $t$, so that
\begin{align}
    x &= \sum_{n=0}^2 a_n t^n,
    &
    y &= \mathhl{\sum_{m=0}^3 b_m t^m}.
\end{align}
This parametrization in the variable $t$ has a symmetry of the form $t \mapsto c_0 + c_1 t$, which has corresponding transformations for the coefficients $a_n$, $b_n$, that keeps the curve in the $x$-$y$ plane unchanged. To remove this symmetry, we select two of the coefficients by choosing $t=0$ to correspond to the extremal value of $x$, $a_1 = 0$, and the rate of change of $y$ around this point to be unity, $b_1=1$.

To fit the curve to the data $(x_i,y_i)$, we recognise that we must not only determine the coefficients $a_n$, $b_n$, but also a set of values $t_i$ which are the parameter values for the closest point on the curve to the data point. We do this employing the residuals
\begin{align}
    \epsilon_i(\bm{\beta}) &= x_i - f_i(\bm{\beta}) 
    &&\text{where}&
    f_i(\bm{\beta}) &= \sum_{n=0}^2 a_n t_i^n,
\end{align}
and
\begin{align}
    \delta_i(\bm{\beta}) &= y_i - g_i(\bm{\beta}) 
    &&\text{where}&
    g_i(\bm{\beta}) &= \sum_{m=0}^3 b_n t_i^m,
\end{align}
where bold characters denote column vectors or matrices, and $\bm{\beta}$ will be defined later. For now we simply state it is the value to be optimised. The sum of squared residuals is then
\begin{align}
    S(\bm{\beta}) &= \sum_i ( \epsilon_i(\bm{\beta})^2 + \delta_i(\bm{\beta})^2 ).
\end{align}
and we employ the method of least squares, seeking a local minimum of $S(\bm{\beta})$. Following the standard deviation for the method of least squares, we can find a local minimum by iterating from one value, $\bm{\beta}^k$, to the next, $\bm{\beta}^{k+1}$, using
\begin{align} \label{eqn:LSQ_itteration}
    (\bm{F}^T\bm{F} + \bm{G}^T\bm{G}) \Delta \bm{\beta}  &= \bm{F}^T \bm{\epsilon} + \bm{G}^T \bm{\delta}
\end{align}
where $\bm{\beta}^{k+1} = \bm{\beta}^{k} + \Delta \bm{\beta}$ and 
\begin{align}
    F_{ij} &= \frac{\partial f_i}{\partial \beta_j} (\bm{\beta}^k),  &
    G_{ij} &= \frac{\partial g_i}{\partial \beta_j} (\bm{\beta}^k).
\end{align}
The algorithm we use proceeds as follows, where $\|\bm{\beta}\| = \frac{1}{M} \sum_{m=1}^M \beta_m$.
\begin{enumerate}
    \item Initialise the values of $t_i$ to $t_i=1$ for $x_i<1$, $y_i>-2$; to $t_i=0$ for $x_i \geq 1$; and to $t_i=-1$ otherwise. Add to these $t_i$ a random amount between $-0.05$ and $0.05$ to desingularize what follows.
    \item Initialise $a_n$ and \hl{$b_m$} to random values between $0$ and $1$, except for $a_1=0$ and $b_1=1$.
    \item Apply the iteration \eqref{eqn:LSQ_itteration} with $\bm{\beta} = (a_0,a_2,b_0)^T$ until $\|\Delta \bm{\beta}\|<10^{-4}$, which is equivalent to fitting $x$ as a quadratic in $y$, 
    \item \label{enu:LSQ_full_itt} Apply the iteration \eqref{eqn:LSQ_itteration} with $\bm{\beta} = (a_0,a_2,b_0,b_2,b_3,t_0,t_1,\ldots)^T$ and $\bm{\beta}^{k+1} = \bm{\beta}^{k} + \frac{1}{10} \Delta \bm{\beta}$ until $\|\Delta \bm{\beta}\|<10^{-4}$, which is the full fitting of the curve
    \item \label{enu:LSQ_global_t} For \hl{each point,} search over values of $t$ to find a $t_i$ that is the global minimizer of $\epsilon_i^2+\delta_i^2$, to ensure that all data points are identified with the correct points on the curve
    \item Apply step \ref{enu:LSQ_full_itt} again
\end{enumerate}

At this stage the optimal values of $a_n$ and $b_m$ are known, and can be used to plot the best fit curve. For both cases $a_1 = 0$ and $b_1 = 1$. For the fluvial case
\begin{align}
    &\begin{aligned}
    \mathhl{a_0} &= \mathhl{0.332},    &
    \mathhl{a_1} &= \mathhl{0},        &
    \mathhl{a_2} &=\mathhl{-1.412},
    \\
    \mathhl{b_0} &= \mathhl{-1.281},   &
    \mathhl{b_1} &= \mathhl{1},        &
    \mathhl{b_2} &= \mathhl{-0.583},   &
    \mathhl{b_3} &= \mathhl{0.1621},
    \end{aligned} \notag
    \\&
    \mathhl{-1.282 < t < 1.179.}
\end{align}
\bibliography{Reference_supp.bib}